\documentclass[twocolumn]{aastex631}
\usepackage{graphicx} 
\usepackage{natbib}
\def\farcs{\hbox{$.\!\!^{\prime\prime}$}} 
\newcommand{\code}[1]{\texttt{#1}}
\usepackage{dcolumn,booktabs}
\usepackage{amsmath}
\usepackage{hyperref}

\usepackage{savesym}
\savesymbol{tablenum}
\usepackage{siunitx}
\restoresymbol{SIX}{tablenum}

\begin{document}
\title{Radial and vertical constraints on the icy origin of H$_2$CO in the HD 163296 Protoplanetary Disk}

\author[0009-0009-2320-7243]{Claudio Hernández-Vera}
\affiliation{Instituto de Astrofísica, Pontificia Universidad Católica de Chile, Av. Vicuña Mackenna 4860, 7820436 Macul, Santiago, Chile}

\author[0000-0003-4784-3040]{Viviana V. Guzmán}
\affiliation{Instituto de Astrofísica, Pontificia Universidad Católica de Chile, Av. Vicuña Mackenna 4860, 7820436 Macul, Santiago, Chile}

\author[0000-0002-8546-9531]{Elizabeth Artur de la Villarmois}
\affiliation{European Southern Observatory, Alonso de Córdova 3107, Casilla 19, Vitacura, Santiago, Chile}

\author[0000-0001-8798-1347]{Karin I. Öberg}
\affiliation{Center for Astrophysics $|$ Harvard \& Smithsonian, 60 Garden St., Cambridge, MA 02138, USA}

\author[0000-0003-2076-8001]{L. Ilsedore Cleeves}
\affiliation{Department of Astronomy, University of Virginia, Charlottesville, VA 22904, USA}

\author[0000-0001-5217-537X]{Michiel R. Hogerheijde}
\affiliation{Leiden Observatory, Leiden University, PO Box 9513, 2300 RA Leiden, The Netherlands}
\affiliation{Anton Pannekoek Institute for Astronomy, University of Amsterdam, PO Box 94249, 1090 GE Amsterdam, The Netherlands}

\author[0000-0001-8642-1786]{Chunhua Qi}
\affiliation{Center for Astrophysics $|$ Harvard \& Smithsonian, 60 Garden St., Cambridge, MA 02138, USA}

\author[0000-0003-2251-0602]{John Carpenter}
\affiliation{Joint ALMA Observatory, Avenida Alonso de Córdova 3107, Vitacura, Santiago, Chile}

\author[0000-0001-8109-5256]{Edith C. Fayolle}
\affiliation{Jet Propulsion Laboratory, California Institute of Technology, 4800 Oak Grove Drive, Pasadena, CA 91109, USA}


\begin{abstract}
H$_2$CO is a small organic molecule widely detected in protoplanetary disks. As a precursor to grain-surface formation of CH$_3$OH, H$_2$CO is considered an important precursor of O-bearing organic molecules that are locked in ices. Still, since gas-phase reactions can also form H$_2$CO, there remains an open question on the channels by which organics form in disks, and how much the grain versus the gas pathways impact the overall organic reservoir. We present spectrally and spatially resolved Atacama Large Millimeter/submillimeter Array observations of several ortho- and para-H$_2$CO transitions toward the bright protoplanetary disk around the Herbig Ae star HD 163296. We derive column density, excitation temperature, and ortho-to-para ratio (OPR) radial profiles for H$_2$CO, as well as disk-averaged values of $N_{\mathrm{T}}\sim4\times 10^{12}$~cm$^{-2}$, $T_{\mathrm{ex}}\sim20$~K, and $\mathrm{OPR}\sim2.7$, respectively. We empirically determine the vertical structure of the emission, finding vertical heights of $z/r\sim0.1$. From the profiles, we find a relatively constant $\mathrm{OPR}\sim2.7$ with radius, but still consistent with $3.0$ among the uncertainties, a secondary increase of $N_{\mathrm{T}}$ in the outer disk, and low $T_{\mathrm{ex}}$ values that decrease with disk radius. Our resulting radial, vertical, and OPR constraints suggest an increased UV penetration beyond the dust millimeter edge, consistent with an icy origin but also with cold gas-phase chemistry. This Herbig disk contrasts previous results for the T Tauri disk, TW Hya, which had a larger contribution from cold gas-phase chemistry. More observations of other sources are needed to disentangle the dominant formation pathway of H$_2$CO in protoplanetary disks. 
\end{abstract}


\section{Introduction}

Planets are formed in disks made of dust and gas around young stars, known as protoplanetary disks. Since part of their composition is thought to be incorporated into nascent planets, they are considered a material reservoir for planetary assembly \citep{Williams2011,Andrews2020,Miotello2023}. Hence, the chemical composition of protoplanetary disks is crucial when disentangling what kind of planets can be formed around other stars \citep[see][for a recent review]{Oberg2023}.

Among the variety of molecules detected in disks, a significant portion are organic species \citep{McGuire2022}, ranging from small organics to complex organic molecules (COMs) composed of six or more atoms \citep{Herbst2009}. COMs have been broadly studied across interstellar and circumstellar environments \citep{Jorgensen2020,Oberg2021,Ceccarelli2023}, as they can potentially be precursors of prebiotic species \citep[e.g.,][]{Garrod2013,Rivilla2019}. Tracing the organic inventory in protoplanetary disks is therefore considered a critical factor in estimating the potential habitability of other planetary systems \citep{Booth2021,Oberg2021} and disentangling the origins of life in our planetary system \citep{Altwegg2019,Ceccarelli2023}.

Although complex hydrocarbons and N-bearing COMs are commonly detected in protoplanetary disks, O-bearing COMs detections are scarce \citep[see][and references therein]{Oberg2023} because most of them are expected to be locked in the ice mantles on the surface of dust grains \citep{Oberg2011,Boogert2015}. Even methanol (CH$_3$OH), one of the simplest O-bearing COMs and considered a building block in forming more complex species \citep{Oberg2009}, has had relatively few detections in disks. Except for TW Hya \citep{Walsh2016}, the closest protoplanetary disk, CH$_3$OH detections are typically restricted to peculiar sources with specific warm conditions \citep[e.g.,][]{vantHoff2018,Lee2019,Podio2020} or cavities directly exposed to the radiation of the host star \citep[e.g.,][]{Booth2021,vanderMarel2021,Booth2023}, consistent with an icy origin. Thus, the need arises for an abundant gas-phase molecule capable of tracing the O-rich organic content in protoplanetary disks.

According to previous work \citep{Walsh2014,Oberg2017}, the small organic molecule formaldehyde (H$_2$CO) has the potential to be a tracer of the frozen content in dust grains since it can efficiently form on ices by the successive hydrogenation of CO, which also leads to the formation of CH$_3$OH and other COMs \citep[e.g.,][]{Watanabe2002,Chuang2017} and is more accessible to detect in the gas-phase due to lower desorption energies compared to CH$_3$OH \citep{Carney2019}. Still, H$_2$CO can also be formed through gas-phase reactions \citep[e.g.,][]{Fockenberg2002,Atkinson2006}, and thus, it cannot be used as a direct tracer of dust-grain chemical products if there are no constraints on the contribution from each formation pathway \citep{Loomis2015,TvS2021}.

Consequently, H$_2$CO has been extensively studied in multiple protoplanetary disks, mostly around low-mass T Tauri stars. As different regions of the disk are expected to have distinct types of chemistry \citep{Oberg2021}, most studies have focused on resolving the radial distribution of H$_2$CO \cite[e.g.,][]{Loomis2015,Oberg2017,Kastner2018,Podio2019,Garufi2020,Guzman2021,Booth2023}. Nevertheless, the disk vertical structure also has an important role in H$_2$CO chemistry, which can be constrained by observing edge-on sources \citep[e.g.,][]{Podio2020}, using geometrical methods \citep[e.g.,][]{Paneque-Carreno2023}, or determining the excitation conditions by observing multiple H$_2$CO lines \citep[e.g.,][]{Pegues2020}. 

\begin{deluxetable}{c c c}
\tabletypesize{\footnotesize}
\tablewidth{0pt} 
\tablenum{1}
\label{table:disk-params}
\tablecaption{Stellar and disk parameters of HD 163296}
\tablehead{
\colhead{Parameter} & \colhead{Value} & \colhead{Reference}}
\startdata
Distance & $101$~pc & \cite{Gaia2018} \\ 
Inclination\tablenotemark{\scriptsize{a}} & $46.7$~deg & \cite{Huang2018} \\
P.A.\tablenotemark{\scriptsize{a}} & $133.3$~deg & \cite{Huang2018} \\
$M_{*}$ & $2.0$~$M_{\sun}$ & \cite{Teague2021} \\
$L_{*}$ & $17.0$~$L_{\sun}$ & \cite{Fairlamb2015} \\
Stellar type & A1 & \cite{Fairlamb2015} \\
$v_{\mathrm{sys}}$ & $5.8$~km~s$^{-1}$ & \cite{Teague2019} \\
\enddata
\tablecomments{Values reproduced from \cite{Oberg2021I}, where additional source properties are included.
\tablenotetext{\scriptsize{a}}{The geometric parameters of the disk are derived from the dust millimeter continuum.}}
\end{deluxetable}

Moreover, H$_2$CO has two isomeric forms depending on the hydrogen nuclei spin alignment: ortho-H$_2$CO (parallel alignment) and para-H$_2$CO (antiparallel alignment). The ratio of the two isomers, known as the ortho-to-para ratio (OPR), has also been considered a possible way to discriminate between H$_2$CO formation pathways \citep[e.g.,][]{Guzman2011}. However, the interpretation of derived OPRs is still a matter of discussion \citep[][see Section~\ref{subsubsec:OPR} for further details]{Guzman2018,TvS2021}. Currently, the excitation conditions and the OPR of H$_2$CO have been well-resolved only for the T Tauri disk TW Hya \citep{TvS2021}, thanks to the Atacama Large Millimeter/submillimeter Array (ALMA) high angular resolution observations of several ortho and para lines of H$_2$CO. Therefore, further multiple-line, spatially resolved studies of p-H$_2$CO and o-H$_2$CO are needed to gain a representative understanding of H$_2$CO chemistry in protoplanetary disks.

Due to its large size \citep[$\sim$$400$~au,][]{Law2021III}, proximity, and high luminosity (see Table~\ref{table:disk-params}), the Herbig Ae disk HD 163296 is an ideal source for a detailed study of H$_2$CO chemistry. Currently, only two p-H$_2$CO transitions have been spatially resolved in HD 163296 \citep{Carney2017,Pegues2020,Guzman2021}. Still, additional bright p-H$_2$CO and o-H$_2$CO lines can be potentially resolved \citep{Guzman2018}, thus covering a broad range of excitation energies to obtain well-constrained estimates of the excitation conditions and the OPR of H$_2$CO. The disk also contains a well-defined CO snowline \citep{Qi2015} and dust millimeter edge \citep{Law2021III}, which makes HD 163296 a perfect target to see how H$_2$CO chemistry works in more massive Herbig Ae/Be sources and compare it to the chemistry in the colder T Tauri, TW Hya disk.

In this work, we analyze the emission of multiple o-H$_2$CO and p-H$_2$CO transitions in HD 163296 resolved at $0\farcs6$ angular resolution with ALMA. We determine empirically the radial and vertical distribution of H$_2$CO, as well as its excitation conditions and OPR as a function of radius. The observations and data reduction are described in Section~\ref{sec:Obs}. The results on the spatial distribution, excitation conditions, and OPR are presented in Section~\ref{sec:Results}, while their implications in the formation of H$_2$CO in protoplanetary disks are discussed in Section~\ref{sec:Discussion}. The conclusions are summarized in Section~\ref{sec:Conclusions}.

\begin{deluxetable*}{c c c c r c c c}
\tabletypesize{\footnotesize}
\tablewidth{0pt} 
\tablenum{2}
\label{table:obs-params}
\tablecaption{Observational parameters of formaldehyde transitions in HD 163296}
\tablehead{
\colhead{H$_2$CO Line} & \colhead{Baselines} & \colhead{Chan. Width} & \colhead{Beam\tablenotemark{\scriptsize{a}}} & \colhead{Beam PA} & \colhead{Chan. rms\tablenotemark{\scriptsize{b}}} & \colhead{Mom. Zero rms\tablenotemark{\scriptsize{c}}} & \colhead{Int. Flux Density\tablenotemark{\scriptsize{d}}}\\
\colhead{} & \colhead{(m)} & \colhead{(km~s$^{-1}$)} & \colhead{($^{\prime\prime}$)} & \colhead{($^\circ$)} & \colhead{(mJy~beam$^{-1}$)} & \colhead{(mJy~beam$^{-1}$ km~s$^{-1}$)} & \colhead{(mJy~km~s$^{-1}$)}}
\decimals
\startdata
\multicolumn{8}{c}{ALMA Band 3} \\
o-H$_2$CO $6_{15}-6_{16}$\tablenotemark{\scriptsize{1}} & $15-3638$ & $0.5$ & $0.6 \times 0.6$ & $-41.20$ & $0.70$ & $0.38$ & $<26$\\ 
\hline 
\multicolumn{8}{c}{ALMA Band 6} \\
p-H$_2$CO $3_{03}-2_{02}$\tablenotemark{\scriptsize{1}} & $15-3638$ & $0.2$ & $0.6 \times 0.6$ & $47.71$ & $1.20$ & $0.48$ & $1011 \pm 13$ \\
o-H$_2$CO $3_{12}-2_{11}$\tablenotemark{\scriptsize{2}} & $15-1039$ & $0.2$ & $0.6 \times 0.6$ & $45.21$ & $1.52$ & $0.54$ & $1595 \pm 13$\\
\hline 
\multicolumn{8}{c}{ALMA Band 7} \\
p-H$_2$CO $4_{04}-3_{03}$\tablenotemark{\scriptsize{2}} & $9-460$ & $0.2$ & $0.6 \times 0.6$ & $-46.72$ & $2.47$ & $0.83$ & $1519 \pm 11$\\
p-H$_2$CO $4_{23}-3_{22}$\tablenotemark{\scriptsize{2}} & $9-460$ & $0.2$ & $0.6 \times 0.6$ & $-71.43$ & $2.51$ & $0.79$ & $137 \pm 11$ \\
p-H$_2$CO $4_{22}-3_{21}$\tablenotemark{\scriptsize{2}} & $9-460$ & $0.2$ & $0.6 \times 0.6$ & $52.56$ & $2.61$ & $0.84$ & $141 \pm 11$\\
o-H$_2$CO $4_{13}-3_{12}$\tablenotemark{\scriptsize{2}} & $9-460$ & $0.2$ & $0.6 \times 0.6$ & $3.84$ & $2.53$ & $0.81$ & $2541 \pm 13$ \\
\enddata
\tablecomments{ALMA project codes: \tablenotemark{\scriptsize{1}} 2018.1.01055.L, \tablenotemark{\scriptsize{2}} 2016.1.00884.S
\tablenotetext{\scriptsize{a}}{The robust parameter used for Briggs weighting and a UV taper, both applied in the CLEAN process, were computed independently for each transition to obtain a common circularized beam of $0\farcs6 \times 0\farcs6$ for all lines (see Appendix~\ref{sec:Appendix_imaging}).}
\tablenotetext{\scriptsize{b}}{The channel rms is calculated using the central $50\%$ of the pixels in the first and last five channels of the data cubes.}
\tablenotetext{\scriptsize{c}}{The moment zero rms is defined by the bootstrapping procedure described in Section~\ref{subsec:spatial-distrib}.}
\tablenotetext{\scriptsize{d}}{The integrated flux density is measured within an outer radius of $4\farcs5$ and integrating the emission between $-1.8$ and $13.4$~km~s$^{-1}$. The uncertainties do not include the $10\%$ calibration uncertainty.}
}
\end{deluxetable*}

\section{Observations}\label{sec:Obs}

\subsection{Observational Details}

We present ALMA observations of several H$_2$CO lines in Bands 3, 6, and 7 in the protoplanetary disk HD 163296\footnote{HD 163296 protoplanetary disk coordinates are $\alpha(2000) = 17^\mathrm{h}56^\mathrm{m}21.280^\mathrm{s}$; $\delta(2000) = -21^\circ57^\prime22.441^{\prime\prime}$.}. A detailed description of the Band 6 and 7 observations (project 2016.1.00884.S, PI: V.V. Guzmán) is given in \cite{Carney2019}. In summary, the Band 6 observations were carried out with the 12 m array in two different configurations between 2016 November and 2017 March, whereas in the case of Band 7, the short baselines executions were observed with the Atacama Compact Array on 2016 October and the long baselines executions were done with the 12m array on 2017 April. Two different spectral setups were used: one targeting the o-H$_2$CO $J=3_{12}-2_{11}$ rotational transition centered at $225.698$~GHz (Band 6), and other targeting several p-H$_2$CO and o-H$_2$CO $J=4-3$ rotational transitions ($J=4_{04}-3_{03}$, $4_{23}-3_{22}$, $4_{22}-3_{21}$, and $4_{13}-3_{12}$) centered at $300.837$~GHz (Band 7). Transitions of other molecules were also covered within the above setups, such as CH$_3$OH \citep{Carney2019} and CN \citep{Bergner2021}.

Additionally, these observations were complemented with publicly available data from the MAPS ALMA Large Program (project 2018.1.01055.L, PI: K.I. Öberg). All the observational details of these data can be found in \cite{Oberg2021I}. In particular, we used the detected p-H$_2$CO $J=3_{03}-2_{02}$ rotational transition at $218.222$~GHz (Band 6) presented by \cite{Guzman2021}, and the nondetected o-H$_2$CO $J=6_{15}-6_{16}$ rotational line at $101.333$~GHz (Band 3). All the transitions used in this work are summarized in Table~\ref{table:obs-params} along with their respective project codes.

\begin{figure*}[ht]
\centering
\includegraphics[width=1.0\linewidth]{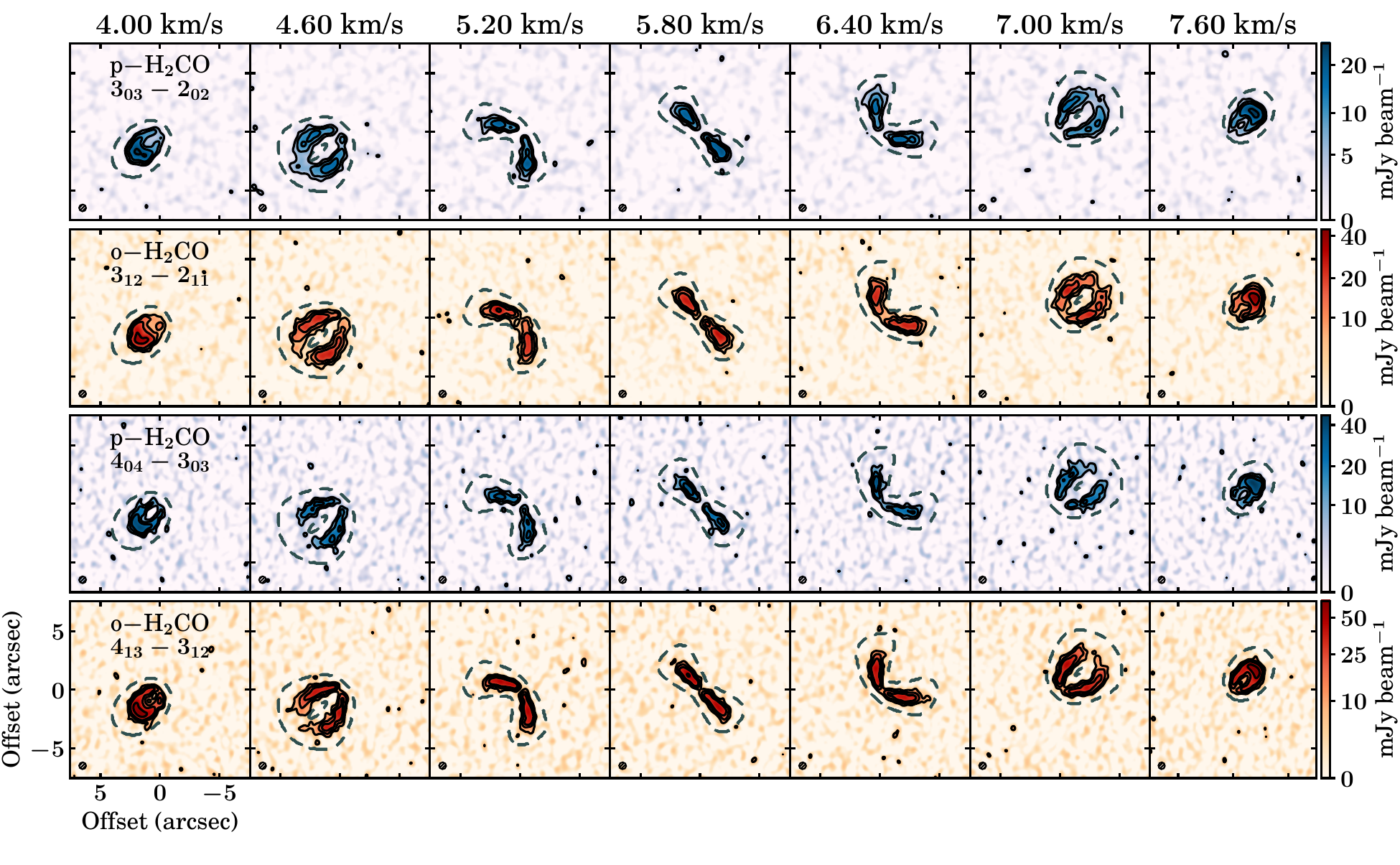}
\caption{Channel maps of the brightest H$_2$CO lines detected toward HD 163296 with the Keplerian mask overlaid (gray-dashed lines). The blue and orange colors represent the emission of the para and ortho isomers, respectively. Each row shows the emission for a particular transition, which is labeled in the leftmost panels, and each column represents a different velocity channel, whose velocities are labeled on top. Contours are $[3,6,10,20]\times\sigma$, where $\sigma$ is the corresponding channel rms reported in Table~\ref{table:obs-params}. The beam size is shown in the bottom left corner of each panel. The maps are shown with an arcsinh color stretch to accentuate the fainter extended emission.}
\label{fig:brightest-lines}
\end{figure*}

\subsection{Data Reduction}

The data reduction and posterior imaging were performed using the Common Astronomy Software Application (\code{CASA})\footnote{https://casa.nrao.edu/} version 6.5.2.26 \citep{McMullin2007,CASA2022}. The data from 2016.1.00884.S were initially processed through the standard pipeline provided by the Joint ALMA Observatory (JAO), and then self-calibrated to improve the signal-to-noise ratio following a methodology similar to that used in the DSHARP ALMA Large Program \citep[see][for more details]{Andrews2018}. Essentially, we created pseudo-continuum visibilities by flagging the line emission in the observed spectral windows, and then we aligned the different measurement sets (short and long baselines separately) into a common phase center by using the \code{CASA} functions \code{fixvis} and \code{fixplanets}. Afterward, the self-calibration procedure can be summarized as follows: we first applied the phase-only self-calibration to the short baseline visibilities, followed by a phase-amplitude self-calibration. Finally, we repeated the same procedures for the combined measurement sets (short + long baselines).

The different calibrations were applied to the data by using the \code{CASA} tasks \code{gaincal} and \code{applycal}. Depending on the band (Band 6 or 7) and the baseline length (only short baselines or combined), we applied between two and seven iterations of phase-only self-calibration using different solution interval steps (``inf," 360, 180, 60, 30, 18, and 6 s) and only one round of phase-amplitude self-calibration using a solution interval of ``inf." The final solutions were then applied to the line data by restoring the flagged spectral line channels, and the continuum was subtracted using the \code{uvcontsub} task. In the case of the MAPS data, we downloaded from the large program webpage\footnote{https://alma-maps.info/data.html} the measurement sets already self-calibrated and continuum-subtracted with the procedure described in \cite{Oberg2021I}.

\begin{figure*}[ht]
\centering
\includegraphics[width=1.0\linewidth]{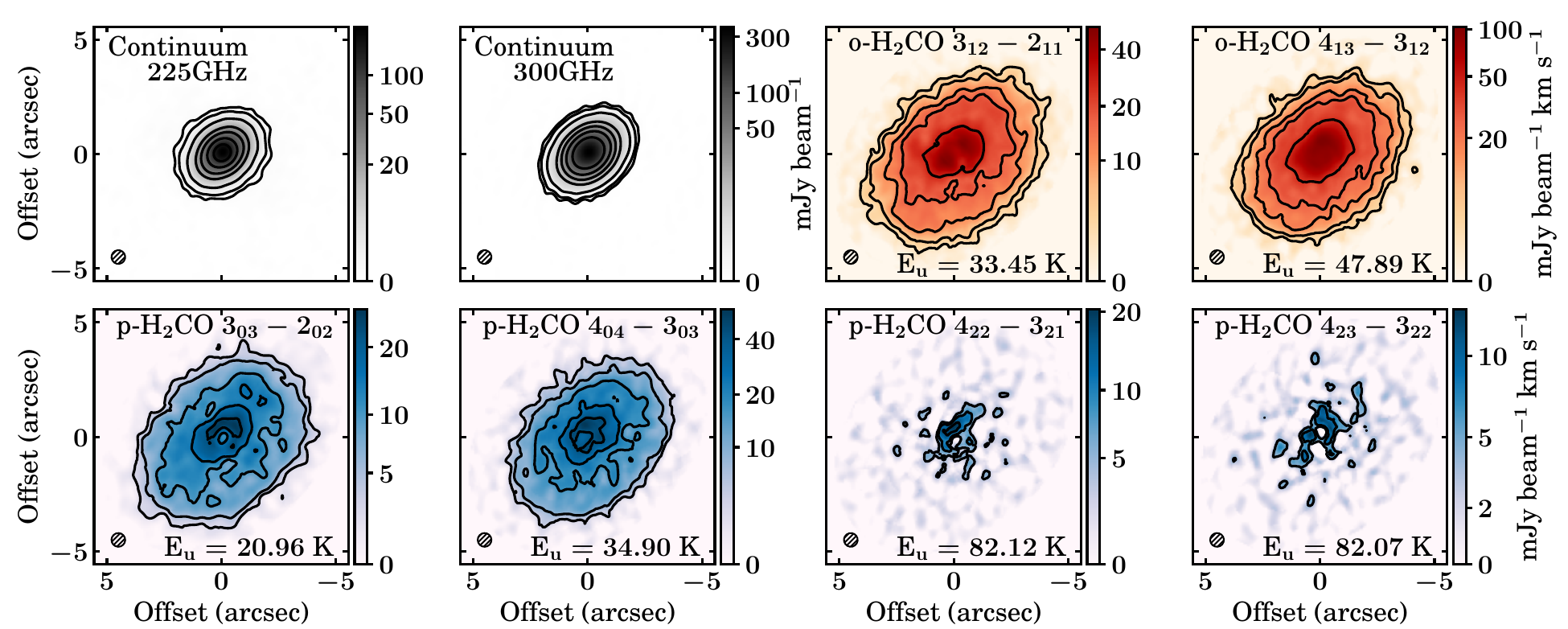}
\caption{Dust continuum emission at 225~GHz and 300~GHz (gray maps), and zeroth-moment maps corresponding to the detected ortho (orange maps) and para (blue maps) H$_2$CO transitions used in this work. Contours are $[5,10,20,30,50]\times\sigma$ in the case of line emission, and $[5,10,30,100,200,400,800]\times\sigma$ for continuum maps, where $\sigma$ is the zeroth-moment rms listed in Table~\ref{table:obs-params} and the continuum rms, respectively. The beam size is shown in the bottom left corner of each panel. The maps are shown with an arcsinh color stretch to accentuate the fainter extended emission.}
\label{fig:h2co-maps}
\end{figure*}

The calibrated visibilities were imaged through the CLEAN algorithm \citep{Hogbom1974} implemented by the \code{tclean} task. We used a Briggs weighting and multiscale deconvolver \citep{Cornwell2008} with scales of 0, 5, 15, and 25 pixels, where the pixel size corresponds to approximately one-seventh of the beam size. The CLEAN mask was defined using the \code{keplerian{\_}mask} package \citep{Teague2020KepMask} to account for the Keplerian rotation of the disk, using the same stellar and disk parameters of HD 163296 as in MAPS \citep[see][values are summarized in Table~\ref{table:disk-params}]{Oberg2021I}, ensuring that we covered all disk emission at different velocities. Regarding continuum images, we used the same elliptical mask as for the self-calibration procedure. Following the same cleaning strategy used in MAPS \citep[see][for a detailed description]{Czekala2021}, we defined the robust parameter jointly with a UV taper of the visibilities in order to obtain a common circularized beam of $0\farcs6 \times 0\farcs6$ for each transition (see Table~\ref{table:imaging_parameters}). The images were CLEANed up to a threshold defined as four times the estimated rms of unmasked regions, and subsequently, were corrected for the \textit{JvM effect} \citep{JvM1995}. The JvM correction is applied by the $\epsilon$ parameter, defined as the ratio between the CLEAN and the dirty beam volumes \citep[see][for more details]{Czekala2021}, whose main effect is a correct accounting of the extended source flux, but which also suppresses the point-source noise in the resulting images. Our derived $\epsilon$ values are listed in Table~\ref{table:imaging_parameters}, where the closer $\epsilon$ is to $1.0$, the smaller its effect. 

The observational parameters of the final JvM-corrected images are summarized in Table~\ref{table:obs-params}. All the final cubes have a channel width of $0.2$~km s$^{-1}$, with the exception of the Band 3 transition, whose native resolution is about $0.5$~km s$^{-1}$. The same procedures where applied then to the continuum measurement sets to obtain the dust continuum images. According to the ALMA Technical Handbook\footnote{https://almascience.eso.org/documents-and-tools/cycle10/alma-technical-handbook}, the reported flux measurements should include an absolute flux calibration uncertainty of approximately $10\%$. 

\section{Results}\label{sec:Results}

\subsection{Disk-integrated Emission}\label{subsec:Disk-int_emission}

Figure~\ref{fig:brightest-lines} shows some representative emission channels of the brightest H$_2$CO transitions and the Keplerian mask used in this work, where we can clearly identify a disk velocity structure consistent with Keplerian rotation. Using the \code{integrated{\_}spectrum} function from the Python package \code{GoFish} \citep{Teague2019GoFish}, we extracted the disk-integrated spectrum within the Keplerian mask for each transition to obtain their corresponding integrated flux densities, which are listed in Table~\ref{table:obs-params}. Following the same bootstrapping procedure described in detail by \cite{Bergner2018} and \cite{Pegues2020}, we defined the uncertainty associated with the integrated flux density as the standard deviation of the integrated fluxes measured within 1000 off-source Keplerian masks.

\begin{figure*}[ht]
\centering
\includegraphics[width=1.0\linewidth]{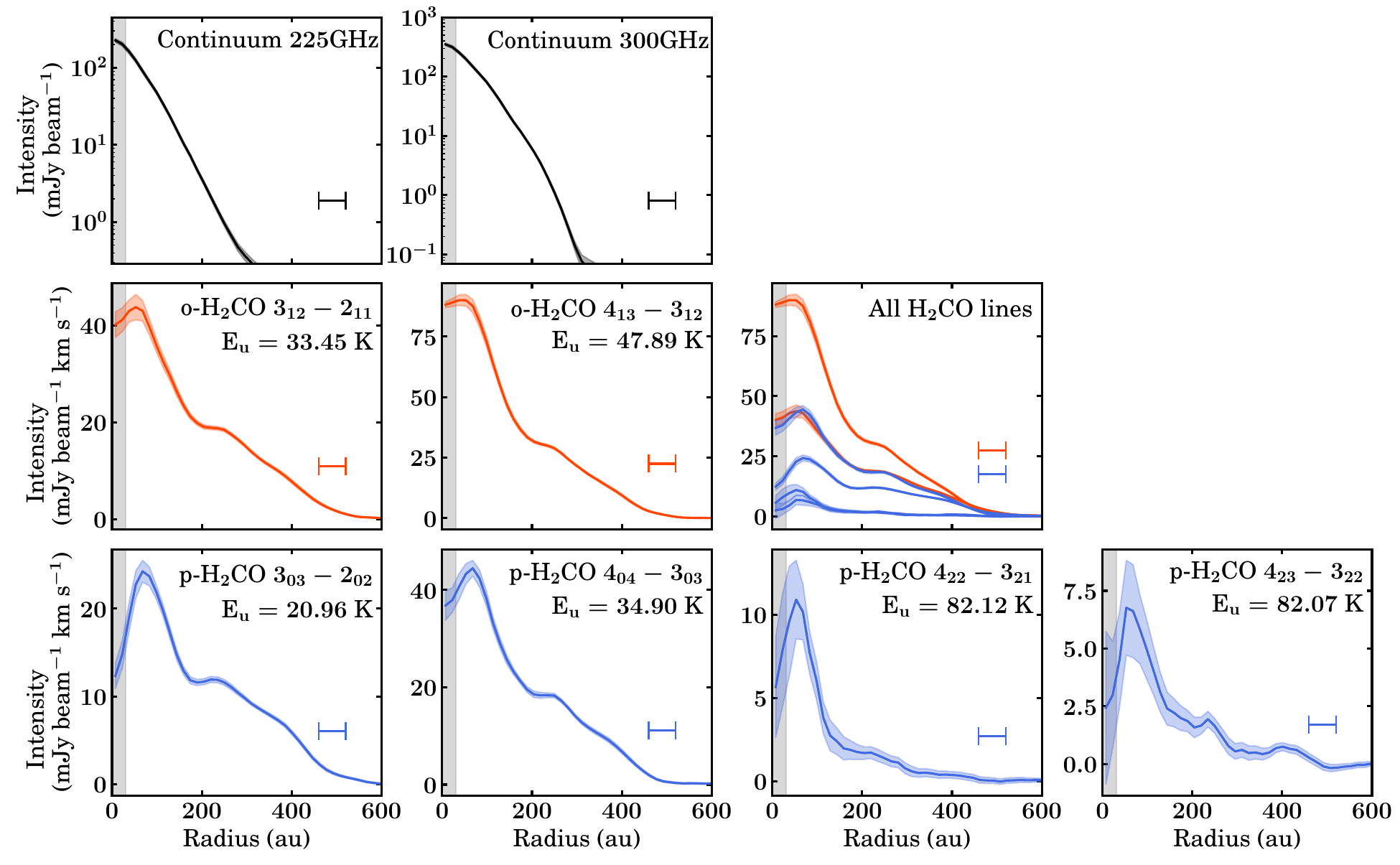}
\caption{Deprojected radial intensity profiles of the dust continuum (top row), o-H$_2$CO (middle row), and p-H$_2$CO (bottom row) emission lines extracted from maps shown in Figure~\ref{fig:h2co-maps}. The color-shaded areas represent the $1\sigma$ scatter, without including the $10\%$ calibration uncertainty. For ease of comparison, all H$_2$CO profiles are shown with the same scale in the rightmost panel of the middle row. The beam size is represented by the horizontal bars, while the vertical gray-shaded regions denote an extent equivalent to half of the beam size, where values should be treated with caution.}
\label{fig:deproj-profiles}
\end{figure*}

\subsection{Spatial Distribution of H$_2$CO Emission}\label{subsec:spatial-distrib}

To analyze the spatial distribution of H$_2$CO in HD 163296, we collapse the spectral cubes into integrated intensity, or \textit{zeroth-moment}, maps using the Python package \code{bettermoments}\footnote{https://github.com/richteague/bettermoments}. The line emission was integrated into velocity using the same Keplerian mask for all transitions (see Figure~\ref{fig:brightest-lines}). We did not apply a threshold clipping to recover accurate fluxes. As the number of selected pixels within a Keplerian mask varies with the different velocity channels, then the associated rms is not uniform across the zeroth-moment map \citep{Bergner2018,Pegues2020}. Thus, we estimated the zeroth-moment rms also using a bootstrapping procedure. We created 1000 zeroth-moment maps, using the same Keplerian mask but centered at different off-source positions. Then, to compute the rms of each pixel, we took the standard deviation of these off-source zeroth-moment maps. Finally, the median value of these standard deviations is considered as the zeroth-moment rms reported in Table~\ref{table:obs-params}.

Figure~\ref{fig:h2co-maps} shows the gallery of zeroth-moment maps, including the dust continuum emission at $225$ and $300$~GHz with the same angular resolution for comparison. The brightest detected lines correspond to transitions with relatively low upper-state energies ($E_u < 50$~K), which are more radially extended than those with higher energies ($E_u \gtrsim 80$~K) and the dust continuum emission. The $6_{15}-6_{16}$ integrated map is not shown since there is no substantial line emission larger than $3\sigma$ neither in the channel maps nor the unmasked zeroth-moment map. Therefore, we use only the six detected transitions for the rest of the analysis.

\subsection{Radial Distribution of H$_2$CO Emission}\label{subsec:radial-distrib}

To have a more precise characterization of the spatial distribution of the emission, we deproject the zeroth-moment maps using the \code{radial{\_}profile} function implemented in \code{GoFish}. The resulting deprojected and azimuthally averaged radial intensity profiles are shown in Figure~\ref{fig:deproj-profiles}, where the continuum profiles are in logarithmic scales to identify the outer edge of the millimeter continuum ($R_{\mathrm{edge}}$) more clearly. We defined $R_{\mathrm{edge}}$ as the radius where the emission falls below $5\sigma$. Considering the two continuum-deprojected radial profiles, we estimated an $R_{\mathrm{edge}}$ value of $\sim$$260$~au, consistent with previous results at higher angular resolution \citep{Carney2017,Law2021III}. On the other hand, the radial profiles of the emission lines are generally more extended than those of the continuum, reaching distances up to $\sim$$450$~au in the case of the four brightest transitions (i.e., $3_{03}-2_{02}$, $3_{12}-2_{11}$, $4_{04}-3_{03}$, and $4_{13}-3_{12}$). 

In terms of morphology, most of the profiles replicate the more prominent radial substructures already witnessed in previous high angular resolution studies of H$_2$CO in HD 163296 \citep[e.g.,][]{Pegues2020,Guzman2021}: a central depression followed by a bright peak at $\sim$$60$~au that then decreases with radius but undergoes two less prominent ring-like structures located at $\sim$$250$~au (spatially coincident with $R_{\mathrm{edge}}$) and $\sim$$380$~au. Regarding continuum morphology, a well-known set of rings and gaps has been previously characterized in the innermost $\sim$$160$~au by higher angular resolution maps \citep[e.g.,][]{Huang2018,Law2021III}, which, unfortunately, are not resolved by our observations.

\subsection{H$_2$CO Excitation Conditions and OPR}\label{subsec:excitation-conditions}

Since we have multiple ortho and para transitions that cover a wide range of upper-state energies ($E_u \sim 20-80$~K), we simultaneously model the emission of each line to determine the excitation temperature, the total (ortho + para) column density, and the OPR of H$_2$CO. Using the Planck function $B_\nu$, the line intensity can be estimated from the excitation temperature, $T_{\mathrm{ex}}$, and the line optical depth, $\tau_\nu$, as follows:

\begin{equation}
\label{eq:I_nu}
    I_{\nu} = (B_\nu(T_{\mathrm{ex}}) - B_\nu(T_{\mathrm{bg}}))[1-e^{-\tau_\nu}],
\end{equation}

\noindent where $T_{\mathrm{bg}}$ is the background temperature defined as the maximum between the dust continuum brightness temperature and the cosmic microwave background (CMB) temperature ($T_{\mathrm{CMB}} = 2.73$~K). 

\begin{deluxetable}{c c c c c c}
\tabletypesize{\footnotesize}
\tablewidth{0pt} 
\tablenum{3}
\label{table:spec-constants}
\tablecaption{Spectroscopic constants of the observed lines}
\tablehead{
\colhead{H$_2$CO Line} & \colhead{Frequency} & \colhead{$\log_{10}(A_{ul})$} & \colhead{$E_u$} & \colhead{$g_u$} & \colhead{$n_{\mathrm{crit}}$}\\
\colhead{} & \colhead{(GHz)} & \colhead{(s$^{-1}$)} & \colhead{(K)} & \colhead{} & \colhead{(cm$^{-3}$)}}
\startdata
\multicolumn{6}{c}{ALMA Band 3} \\
(o) $6_{15}-6_{16}$ & $101.333$ & $-5.80408$ & $87.56$ & $13$ & $3.93\times10^4$\\ 
\hline 
\multicolumn{6}{c}{ALMA Band 6} \\
(p) $3_{03}-2_{02}$ & $218.222$ & $-3.55037$ & $20.96$ & $7$ & $2.56\times10^6$\\
(o) $3_{12}-2_{11}$ & $225.698$ & $-3.55754$ & $33.45$ & $7$ & $4.47\times10^6$\\
\hline 
\multicolumn{6}{c}{ALMA Band 7} \\
(p) $4_{04}-3_{03}$ & $290.623$ & $-3.16132$ & $34.90$ & $9$ & $5.75\times10^6$\\
(p) $4_{23}-3_{22}$ & $291.238$ & $-3.28334$ & $82.07$ & $9$ & $1.02\times10^7$\\
(p) $4_{22}-3_{21}$ & $291.948$ & $-3.28024$ & $82.12$ & $9$ & $9.37\times10^6$\\
(o) $4_{13}-3_{12}$ & $300.837$ & $-3.14420$ & $47.89$ & $9$ & $8.75\times10^6$\\
\enddata
\tablecomments{The rest frequency and Einstein $A$ coefficient are taken from the JPL database \citep{Pickett1998}, while the upper-state energy, the upper-state degeneracy, and the collisional coefficients to compute the critical densities are extracted from the Leiden Atomic and Molecular Database \citep[LAMDA;][]{Schöier2005}.
}
\end{deluxetable}

\begin{figure*}[ht]
\centering
\includegraphics[width=1.0\linewidth]{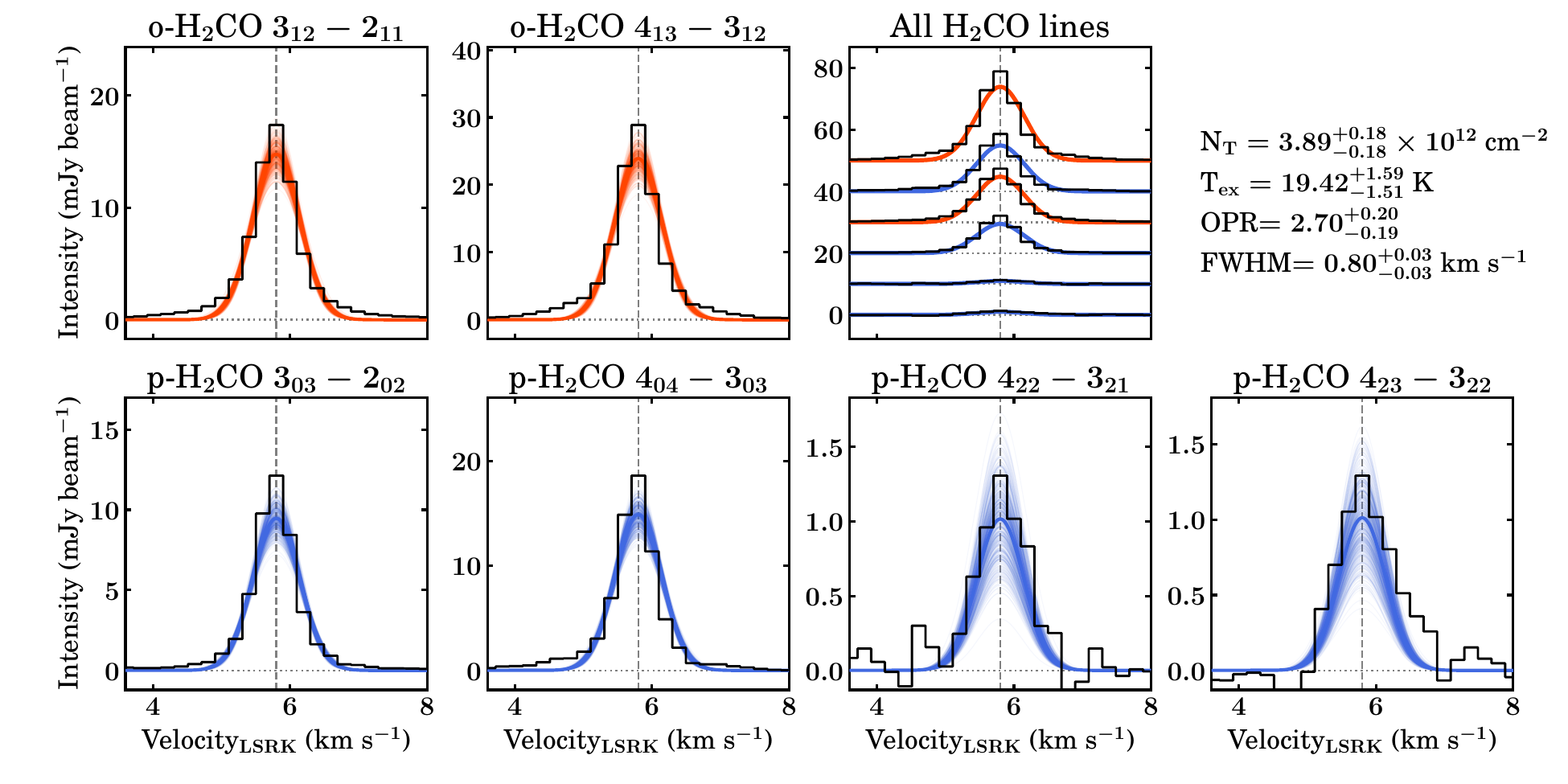}
\caption{Disk-averaged spectra, corrected by the Keplerian rotation, and Gaussian models of all transitions used to fit the excitation conditions of H$_2$CO. The best-fit parameters are listed on the top right side, whereas random draws from the fit posterior distributions are shown in orange and blue for the ortho and para isomers, respectively. For ease of comparison, the best-fit models for all H$_2$CO transitions are shown with the same scale in the rightmost panel of the top row, using a constant vertical shift of $10$ mJy beam$^{-1}$. The corresponding transitions are labeled on top of each panel. The dashed vertical line denotes the systemic velocity of HD 163296 in the kinematic local standard of rest (LSRK) frame.}
\label{fig:fitted_disk-averaged_spectra}
\end{figure*}

At the velocity center of the line, $\tau_\nu$ is defined as

\begin{equation}
\label{eq:tau}
    \tau_0 = \frac{A_{ul}c^3 N_{u}}{8\pi\nu^3\Delta v}(e^{h\nu/kT_{\mathrm{ex}}} - 1),
\end{equation}

\noindent where $\nu$ is the rest frequency of the line, $A_{ul}$ is the Einstein coefficient, $\Delta v$ is the velocity line width define as the FWHM of the line profile, and $N_u$ is the column density of the molecule in the upper state of the transition. If we assume that the line emission occurs under local thermodynamic equilibrium (LTE), then

\begin{equation}
\label{eq:Boltzmann-eq}
    N_u = \frac{N_{\mathrm{T}}}{Q(T_{\mathrm{ex}})}g_u e^{-E_u/kT_{\mathrm{ex}}},
\end{equation}

\noindent where $N_{\mathrm{T}}$ is the total column density, $Q$ is the partition function of H$_2$CO, and $E_u$ with $g_u$ are the energy and the degeneracy associated to the upper-state level of each transition, respectively (see Table~\ref{table:spec-constants}). This should be a reasonable assumption for HD 163296 since the critical densities of the observed transitions ($n_{\mathrm{crit}} \lesssim 10^{7}$~cm$^{-3}$, see Table~\ref{table:spec-constants}) are lower than the typical gas densities estimated by models \citep{Qi2011,Zhang2021} at those vertical heights where we expect to find H$_2$CO \citep[e.g.,][]{Paneque-Carreno2023}. In that sense, the excitation temperature can be used as a direct tracer of the kinetic gas temperature in those disk layers where H$_2$CO is emitting.

\begin{figure*}[ht]
\centering
\includegraphics[width=1.0\linewidth]{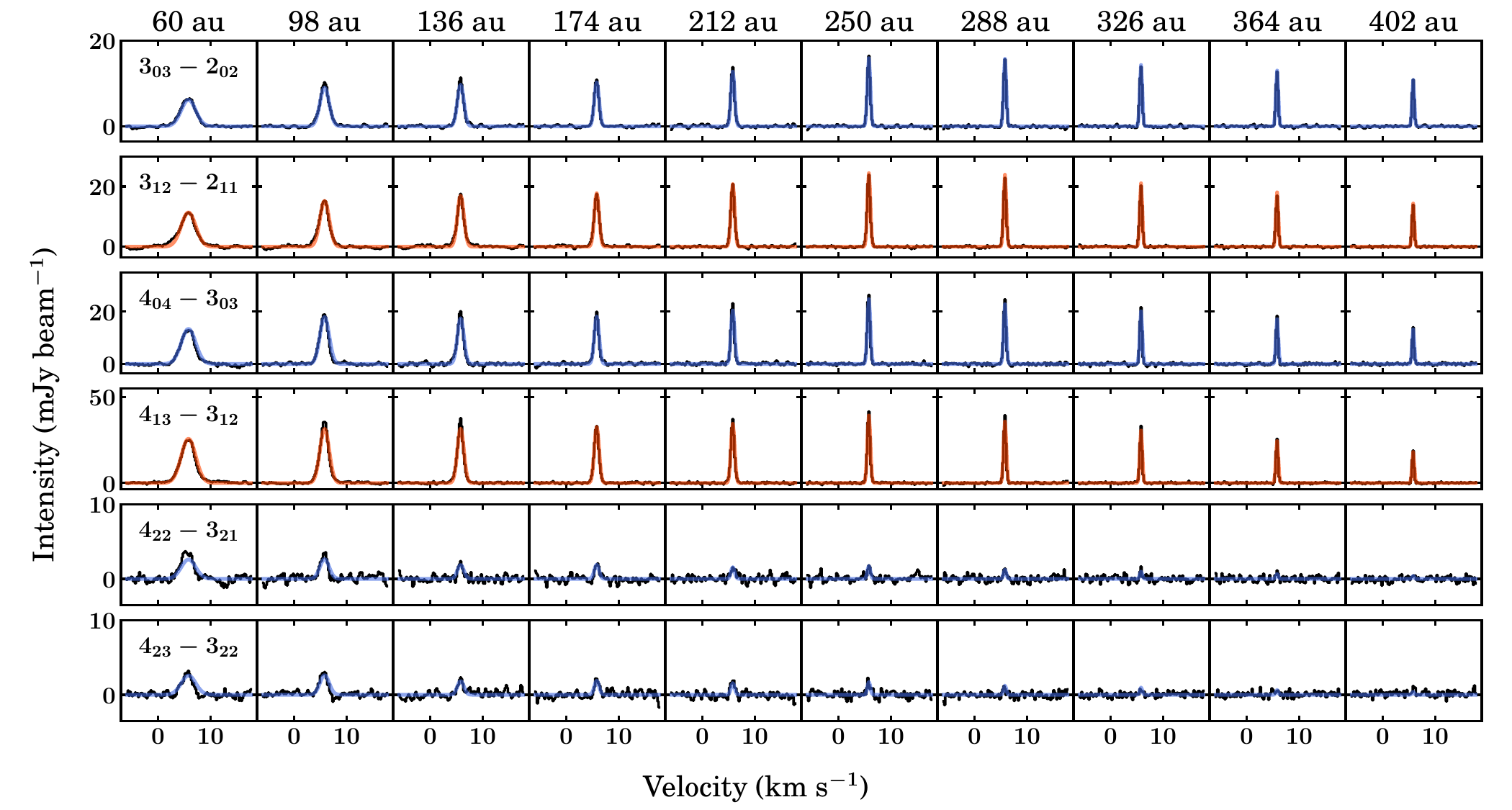}
\caption{Same as Figure~\ref{fig:fitted_disk-averaged_spectra}, but using the deprojected spectra averaged over radial annuli instead of the entire disk-averaged spectra. The bin center of each radial annulus is labeled on top. Best-fit models are shown in orange and blue for the ortho and para isomers, respectively.}
\label{fig:fitted_radial_spectra}
\end{figure*}

We defined separate partition functions for the ortho and para isomers. Using the same convention adopted by \cite{TvS2021}, we defined

\begin{eqnarray}
\label{eq:Qortho}
    Q(T_{\mathrm{rot}})^{\mathrm{ortho}} & = & \sum_i^{\mathrm{ortho}} g_i e^{-E_i/kT_{\mathrm{rot}}}, \: \text{and}\\
\label{eq:Qpara}
    Q(T_{\mathrm{rot}})^{\mathrm{para}} & = & \sum_i^{\mathrm{para}} g_i e^{-E_i/kT_{\mathrm{rot}}},
\end{eqnarray}

\noindent where $g_i$ and $E_i$ are the upper-state degeneracy and energy of a state. The sum in Equations~(\ref{eq:Qortho})~and~(\ref{eq:Qpara}) is performed over the rotational ground states of a specific isomer (ortho or para) taken from the ExoMol database \citep{Al-Refaie2015, Wang2020} but using state degeneracies consistent with the ones from LAMDA \citep{Schöier2005} since the former assumes an OPR of $3.0$ \citep[see][for more details]{TvS2021}.

By putting Equations (\ref{eq:tau}) and (\ref{eq:Boltzmann-eq}) into Equation (\ref{eq:I_nu}), and assuming a Gaussian line profile centered at $v_{\mathrm{sys}}=5.8$~km~s$^{-1}$ \citep{Teague2019} with a certain $\Delta v$, we can model the line intensity as a function of $T_{\mathrm{ex}}$ and $N_{\mathrm{T}}$. As we have different partition functions, and $T_\mathrm{rot}^{\mathrm{ortho}}$ is not necessarily equal to $T_\mathrm{rot}^{\mathrm{para}}$, we could in principle do a separate line modeling for each isomer and then obtain the ortho-to-para ratio $\mathrm{OPR} = N_{\mathrm{T}}^{\mathrm{ortho}}/N_{\mathrm{T}}^{\mathrm{para}}$. However, in this work, we assume the same excitation temperature since we detected only two o-H$_2$CO lines. Hence, we fitted simultaneously the detected ortho and para transitions using the following relations:

\begin{eqnarray}
    N_{\mathrm{T}}^{\mathrm{ortho}} = \frac{N_{\mathrm{T}}\times\mathrm{OPR}}{1+\mathrm{OPR}}, \: \text{and}\: N_{\mathrm{T}}^{\mathrm{para}} = \frac{N_{\mathrm{T}}}{1+\mathrm{OPR}},
\end{eqnarray}

\noindent where $N_{\mathrm{T}} = N_{\mathrm{T}}^{\mathrm{ortho}} + N_{\mathrm{T}}^{\mathrm{para}}$ is the total column density of H$_2$CO considering both isomers. Thus, the free parameters to fit are $N_{\mathrm{T}}$, $T_\mathrm{rot}$, $\mathrm{OPR}$, and $\Delta v$. To fit the excitation conditions and the line width, we construct a likelihood function from Equation~(\ref{eq:I_nu}) and use the \code{emcee} package \citep{Foreman-Mackey2013} to explore the parameter space and get the posterior probability distributions.

\subsubsection{Disk-averaged Analysis}

We first model the line emission averaged over the full radial extent of H$_2$CO ($\sim$$450$~au) in order to derive the disk-averaged excitation conditions. To account for the velocity structure of the disk, we used the \code{average{\_}spectrum} function from \code{GoFish}, which corrects the spectrum for Keplerian rotation before averaging by using the stellar mass and the distance to the disk. For simplicity, we assume that the emitting layer is in the midplane ($z/r = 0$), which should be reasonable based on the estimates of the vertical height of H$_2$CO in HD 163296 (see Section~\ref{subsec:vertical-distrib} for further details). The retrieved disk-averaged spectra are shown in Figure~\ref{fig:fitted_disk-averaged_spectra}, where the best-fit Gaussian models are overlaid with different colors for each H$_2$CO isomer. The best-fit excitation conditions and their corresponding uncertainties are deduced from the 16th, 50th, and 84th percentiles of the posterior distributions, whose results are $N_{\mathrm{T}} = 3.89_{-0.18}^{+0.18} \times 10^{12}$~cm$^{-2}$, $T_{\mathrm{ex}} = 19.42_{-1.51}^{+1.59}$~K, and $\mathrm{OPR} = 2.70_{-0.19}^{+0.20}$ for a Gaussian line width of $\Delta v = 0.80_{-0.03}^{+0.03}$~km~s$^{-1}$. The disk-averaged line optical depths are between $\tau_\nu = 0.003$ and $0.074$, verifying an optically thin regime. We also checked that these excitation conditions are consistent with the nondetection of the $6_{15}-6_{16}$ line, as its predicted line intensity is lower than the sensitivity of the observations.

As shown in Figure~\ref{fig:fitted_disk-averaged_spectra}, the LTE predictions underestimate the disk-averaged emission from the line wings, indicating that the line profiles are probably more Lorentzian than pure Gaussian. Consequently, we performed the same analysis but using a Voigt profile. We find an improvement in the fit quality for the line wings, and recover slightly higher values for $N_{\mathrm{T}}$ and the OPR, although consistent with the results from the Gaussian fitting when considering the uncertainties (see Appendix~\ref{sec:Appendix_voigt}). A similar line profile behavior was found in TW Hya using $^{12}$CO $J=3-2$ line observations, that were attributed to pressure broadening \citep{Yoshida2022}. Our H$_2$CO results may hint at something similar, but further analysis is needed to investigate the origin of the non-Gaussian profiles in HD 163296.

Previous efforts have been made to constrain the distribution of H$_2$CO across HD 163296. For the disk-integrated column density, \cite{Carney2019} estimated a value of about $N_{\mathrm{T}} = 2.1 \times 10^{12}$~cm$^{-2}$, which is lower by a factor of $\sim$$2$ than our derived value. However, it is still consistent, considering that they assumed a higher excitation temperature of $T_{\mathrm{ex}} = 25$~K. Moreover, to compute the total column density, they used integrated flux measurements of a single line from Submillimeter Array (SMA) observations with lower angular resolution \citep{Qi2013}. Thus, the differences between the disk-averaged $N_{\mathrm{T}}$ estimates could be due to choosing a different area of integration, or filtering out more extended emission (either in ALMA or the SMA).

Regarding the excitation temperature, \cite{Carney2017} estimated a lower limit of $>$$20$~K on the kinetic temperature using three p-H$_2$CO $J=3-2$ rotational transitions. Nevertheless, as they did not have higher energy transitions (e.g., $J=4-3$), and their two weakest lines were detected only through a matched filter analysis \citep[see][for details of this technique]{Loomis2018matchedfilter}, the temperature estimation was uncertain. Later, a similar result was found by \cite{Guzman2018}, where a rotational temperature of $T_{\mathrm{rot}} = 24.3 \pm 13.5$~K was obtained using three o-H$_2$CO transitions with lower angular resolution SMA data, consistent with our results considering the uncertainties.

In the case of the ortho-to-para ratio, the only known determination for HD 163296 was also made by \cite{Guzman2018}, who complemented the three o-H$_2$CO transitions mentioned above with three p-H$_2$CO transitions and estimated an ortho-to-para ratio between $\mathrm{OPR} = 1.8$ and $2.8$, depending on the vertical height of the emission. This range of values agrees with our OPR determination. However, since the derived uncertainties are purely statistical and do not consider the assumptions mentioned in Section~\ref{subsec:excitation-conditions}, we cannot rule out a disk-averaged OPR value consistent with $3.0$ in HD 163296 (see Section~\ref{subsubsec:OPR}).

\subsubsection{Radially Resolved Analysis} \label{subsubsec:Resolved_analysis}

\begin{figure}[t]
\centering
\includegraphics[width=1.0\linewidth]{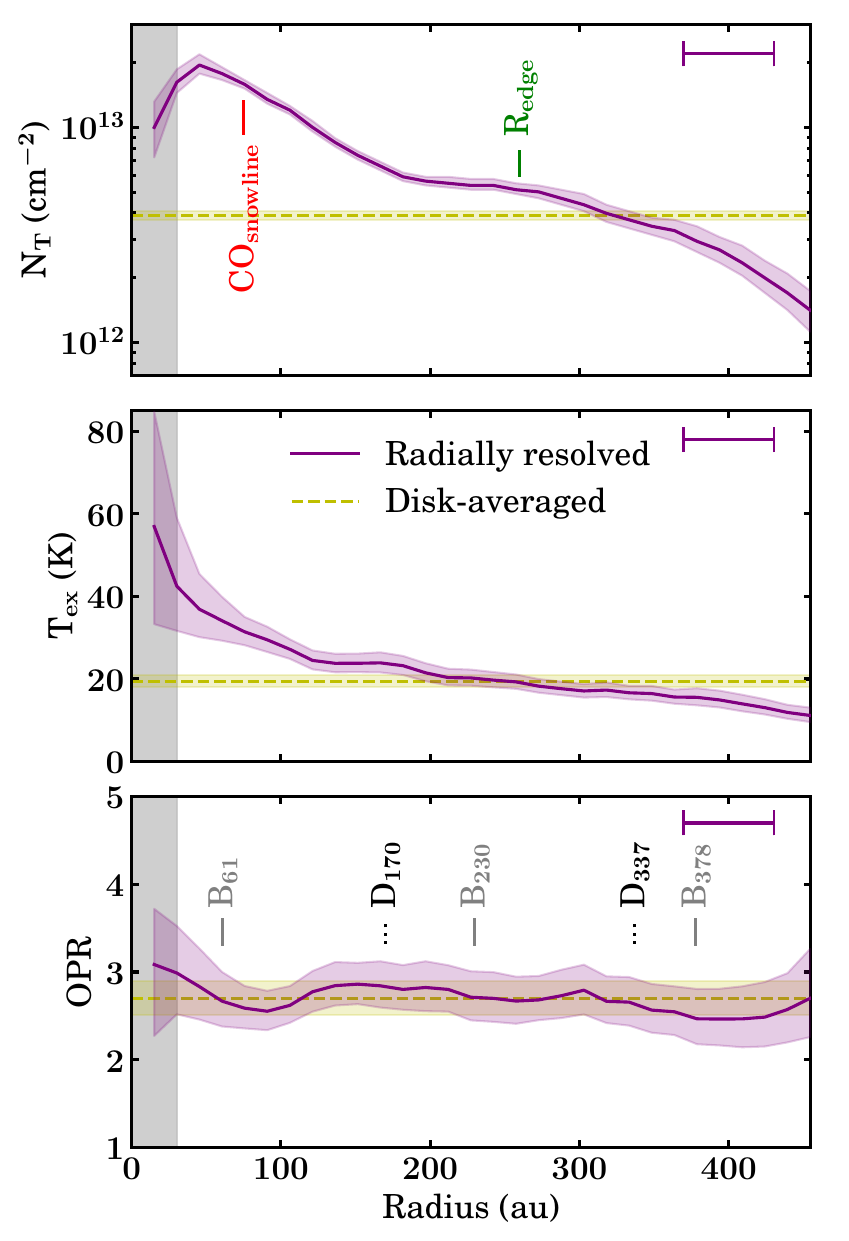}
\caption{Radial profiles of the total column density (top panel), excitation temperature (middle panel), and the ortho-to-para ratio (bottom panel) of H$_2$CO. The solid purple line and the dashed yellow line represent the best-fit values for the radially resolved and disk-averaged analysis, respectively, whose uncertainties correspond to the color-shaded areas. The horizontal bars and the vertical gray-shaded regions are the same as in Figure~\ref{fig:deproj-profiles}. In the top panel, vertical lines depict the CO snowline (red) and the dust millimeter edge (green), whereas in the bottom panel, they indicate those H$_2$CO emission rings (gray) and gaps (black) from \cite{Law2021III} that are resolved with our resolution.}
\label{fig:excitation-conds}
\end{figure}

Since we can resolve the variation of the H$_2$CO emission with the disk radius (see Figure~\ref{fig:deproj-profiles}), we repeat the same line fitting procedure but use the deprojected radial spectra instead of the disk-averaged one to derive the excitation conditions as a function of radius. Using \code{GoFish}, we applied the \code{radial{\_}spectra} function to deproject the spectra into different radial bins. Similarly to the disk-averaged case, the emission is shifted and stacked according to the Keplerian motion. However, it is now azimuthally averaged over different radial annuli whose width corresponds to one-quarter of the beam's major axis. In this case, we construct a likelihood function for each radial bin to determine the best-fit models at different radii, as shown in Figure~\ref{fig:fitted_radial_spectra}.

Figure~\ref{fig:excitation-conds} shows the resulting deprojected radial profiles of the excitation conditions and the disk-averaged values for comparison, whereas $\tau_{\nu}$ radial profiles are shown in Appendix~\ref{sec:Appendix_tauprofiles}. In general, the mean values of each profile are compatible with the disk-averaged results, and we confirmed that the emission is optically thin throughout the disk (see Figure~\ref{fig:tau_profiles}). The derived $N_{\mathrm{T}}$ profile behaves similarly to those from previous studies where the emission of H$_2$CO was resolved in HD 163296 \citep{Pegues2020,Guzman2021}. Total column density values range from $\sim 2.0 \times 10^{13} - 1.3\times 10^{12}$~cm$^{-2}$ within a radius of $\sim450$~au, displaying two remarkable features: a central gap followed by a global peak close to the CO snowline located at $75$~au \citep[][corrected by Gaia distance]{Qi2015} and a secondary ring or plateau near $R_{\mathrm{edge}}\sim 260$~au, as defined by our continuum observations.

Regarding the excitation temperature, this is the first time it has been resolved for H$_2$CO in HD 163296 since, previously, it was computed through a disk-averaged analysis \citep{Carney2017,Guzman2018} or assumed uniform across the disk \citep{Pegues2020,Guzman2021} either due to a lack of sufficient angular resolution or multiple-line observations. We find that the $T_{\mathrm{ex}}$ profile decreases smoothly from $50$ to $10$~K, where a temperature of $\sim$$30$~K is reached close to CO snowline and $\sim$$20$~K at $R_{\mathrm{edge}}$.

On the other hand, the ortho-to-para ratio profile is relatively constant across the disk, mostly centered at $\mathrm{OPR}\sim 2.7$ but with uncertainties that remain compatible with the statistically expected value of $\mathrm{OPR}=3.0$. Nevertheless, the profile exhibits small modulations that are apparently consistent with some of the H$_2$CO radial substructures witnessed by \cite{Law2021III} that are spatially resolved by our observations, which could be linked with changes in the chemistry.

\subsection{Vertical Height of the Emission}\label{subsec:vertical-distrib}

Although the radial distribution of the excitation conditions gives us valuable clues about the formation of H$_2$CO in protoplanetary disks, the latter has a three-dimensional structure where the vertical distribution of the emission is also an essential factor to having a complete understanding of the contribution of the different formation pathways. To complement our radially resolved analysis, we used the Python package \code{ALFAHOR}\footnote{https://github.com/teresapaz/alfahor} to constrain empirically the vertical height of H$_2$CO in HD 163296 from the observations \citep{Paneque-Carreno2023}.

By having the stellar and geometrical properties of the disk, \code{ALFAHOR} computes a vertical profile based on a geometrical method \citep{Pinte2018} that relies on the location of the emission maxima inside a user-defined mask for each channel \citep[see][for more details]{Paneque-Carreno2023}. Using the stellar and disk parameters from Table~\ref{table:disk-params}, and masking manually the channel emission for each H$_2$CO transition, we derived the vertical profiles shown in Figure~\ref{fig:vertical_height}. Since this empirical method requires an appropriate combination of angular resolution and signal-to-noise ratio to distinguish the near and far \textit{wings} of the channel emission \cite[see Figure~1 in][]{Paneque-Carreno2023}, we applied this analysis only to the four brightest transitions in our sample.

H$_2$CO is apparently tracing an emission surface close to $z/r \sim 0.1$. Still, as the lines used in this work have $\tau_\nu \ll 1$, we cannot discard the presence of H$_2$CO at higher disk layers since optically thin emission makes it challenging to isolate surfaces due to back- and front-side confusion across the disk. Thus, the determined value should be interpreted as a lower limit. We did not notice remarkable differences in the height predicted from different lines, which agrees with the LTE assumption where all transitions have the same excitation temperature since the disk gas temperature changes with the radius and the height. Similar to the integrated intensities (see Figure~\ref{fig:deproj-profiles}), the vertical profiles present radial modulations, as previously reported by \cite{Paneque-Carreno2023}, where the lowest heights are spatially coincident with the H$_2$CO rings. This suggests that the structures seen from the radial profiles are likely related to changes in the H$_2$CO chemistry, which could be related to the slight radial variations of the OPR (see Section~\ref{subsec:H2CO_HD163296} for further discussion).

\section{Discussion}\label{sec:Discussion}

\subsection{The chemistry of H$_2$CO in Protoplanetary Disks} \label{H2CO_pathways}

The reliability of H$_2$CO as a tracer of the icy O reservoir in protoplanetary disks will depend on its dominant formation pathway since, at typical disk conditions, it could be formed by gas-phase and grain-surface chemistry \citep{Walsh2014}. According to laboratory experiments \citep[e.g.,][]{Fockenberg2002,Atkinson2006} and chemical modeling \citep[e.g.,][]{vanderMarel2014,Loomis2015}, H$_2$CO is formed efficiently in the gas phase at low temperatures ($<$$100$~K) by the following neutral-neutral reaction:

\begin{equation}
    \mathrm{CH}_3 + \mathrm{O} \rightarrow \mathrm{H}_2\mathrm{CO} + \mathrm{H},
\end{equation}

\noindent which is expected to dominate in warm environments such as the inner disk \citep{Loomis2015,Oberg2017} or surface layers away from the midplane at relatively large normalized heights \citep[$z/r \geq 0.25$,][]{TvS2021}.

On the other hand, in those cold regions where most of the O-containing carriers are frozen, H$_2$CO and other more complex species (e.g., CH$_3$OH) are expected to form by the successive hydrogenation of CO on dust grains \citep[e.g.,][]{Watanabe2002,Fuchs2009}:

\begin{equation}
\label{eq:CO-hydrogenation}
    \mathrm{CO} \xrightarrow{\mathrm{H}} \mathrm{HCO} \xrightarrow{\mathrm{H}} \mathrm{H}_2\mathrm{CO} \xrightarrow{\mathrm{H}} \mathrm{H}_2\mathrm{COH} \xrightarrow{\mathrm{H}}\mathrm{CH}_3\mathrm{OH},
\end{equation}

\noindent which should be relevant in the vicinity and inside the regions with abundant CO ices, delimited by the CO snowline and the CO snow surface \citep{Oberg2017,Guzman2021}. Since thermal desorption of H$_2$CO occurs at warmer conditions than the CO freeze-out temperature \citep{Pegues2020,TvS2021}, nonthermal processes are required to desorb the grain-surface H$_2$CO into the gas-phase.

Based on the above, the key takeaway from previous works is that probably a combination of gas-phase and grain-surface formation pathways is needed to explain the emission morphologies deduced from H$_2$CO observations \cite[e.g.,][]{Pegues2020,Guzman2021}, which is supported by predictions from chemical models that include both kinds of chemistry \citep{Loomis2015,Oberg2017} but also by empirical measurements of vertical stratification in disks \citep{Podio2020,Paneque-Carreno2023}. Still, most previous studies lack multiple transitions and different isomer observations, and therefore, the excitation conditions with the OPR, which are directly related to H$_2$CO formation, are barely constrained.

In fact, there are only two disks where the above quantities are now well-resolved. \cite{TvS2021} recently resolved the excitation conditions and the OPR of H$_2$CO in the protoplanetary disk around TW Hya. They found that H$_2$CO emits from a relatively warm ($30-40$~K) layer, and a decreasing OPR with radius, going from values consistent with 3 within 60~au (which corresponds to the extent of the pebble disk) and values of 2 beyond this radius. The relatively uniform excitation conditions and OPR gradient suggest a single gas-phase formation pathway dominates the formation of H$_2$CO across the TW~Hya disk, although a contribution from grain-surface chemistry cannot be completely ruled out \citep{TvS2021}. The other case is HD 163296, with the observations presented in this work.

\begin{figure*}[ht]
\centering
\includegraphics[width=0.8\linewidth]{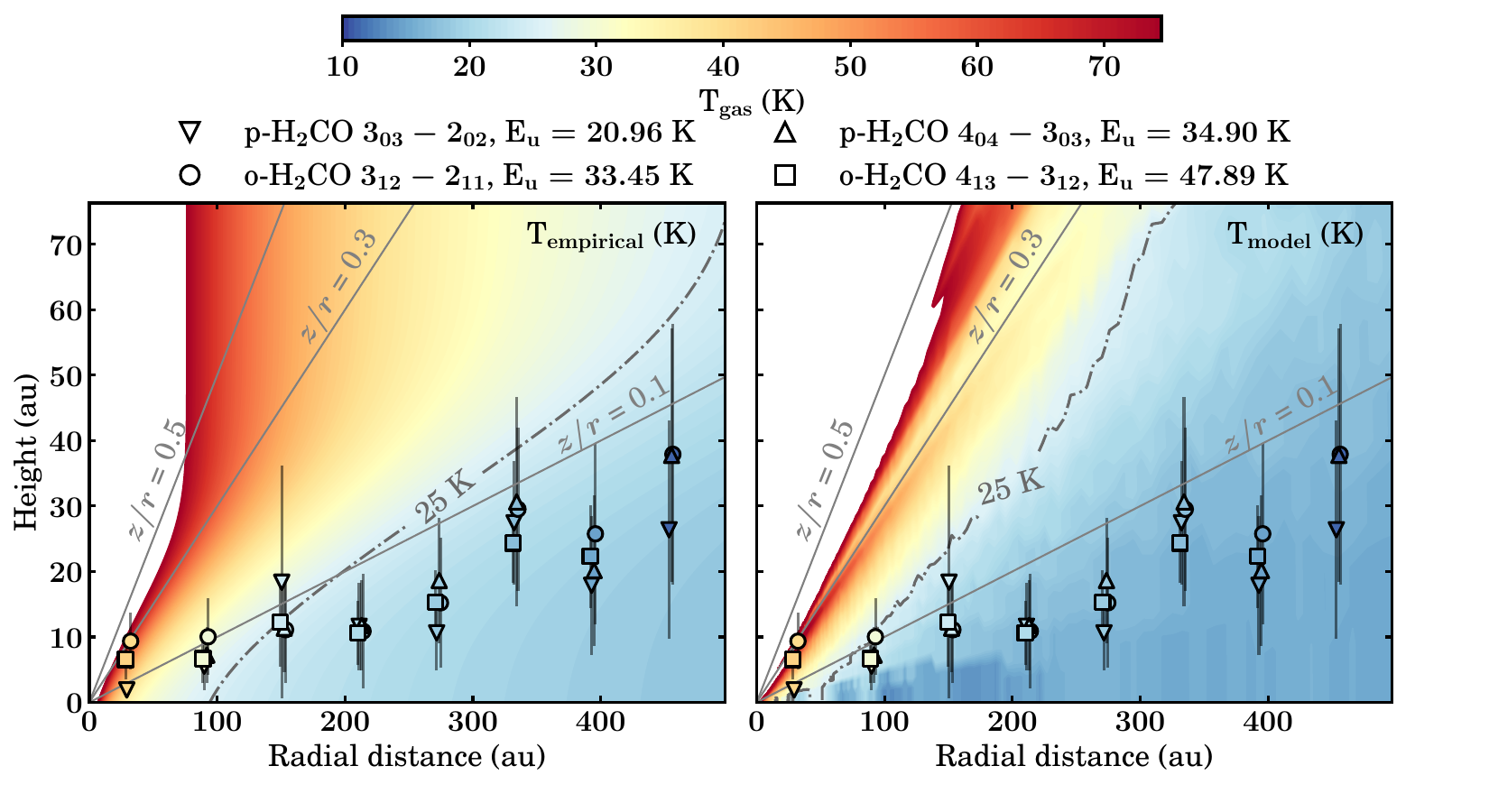}
\caption{Vertical emission profiles of the brightest H$_2$CO lines detected toward HD 163296. Symbols are colored according to the fitted excitation temperature profile (see the middle panel in Figure~\ref{fig:excitation-conds}), and each marker represents a different transition, which is labeled on top. For comparison, the background shows the temperature structure predicted by empirical observations \citep[][left panel]{Law2021IV} and by thermochemical models \citep[][right panel]{Zhang2021}. Solid gray lines represent emission surfaces with constant $z/r=0.1,\:0.3,\:\mathrm{and}\:0.5$ values, whereas dash-dotted gray lines delineate the CO snow surface traced by the CO freeze-out midplane temperature \citep[$\sim$$25$~K,][]{Qi2015}.}
\label{fig:vertical_height}
\end{figure*}

\subsection{Constraints on H$_2$CO Formation in HD 163296} \label{subsec:H2CO_HD163296} 

\subsubsection{Radial Distribution of H$_2$CO} \label{subsubsec:Rad_distrib}

From Figure~\ref{fig:excitation-conds}, we deduce a clear correlation between the substructures of the $N_{\mathrm{T}}$ profile with the CO snowline and $R_{\mathrm{edge}}$, which is likely connected with changes in the chemistry. The same spatial coincidence observed in previous work of H$_2$CO toward HD 163296 \citep{Carney2017,Guzman2021} and other disks \citep[e.g.,][]{Oberg2017,Kastner2018,Pegues2020} has been interpreted as an indicator of active H$_2$CO grain-surface production. Nevertheless, although less efficient at larger radii, the contribution from the gas-phase formation route may not be negligible. 

Despite warm temperatures ($T_{\mathrm{ex}} \gtrsim 30$~K, see middle panel in Figure~\ref{fig:excitation-conds}), there is a substantial amount of H$_2$CO in the inner disk ($\lesssim$$100-150$~au). Particularly, within the CO snowline, we would not expect to have the formation of abundant CO ices that could form H$_2$CO on dust grains. This warm H$_2$CO component is generally ascribed to warm gas-phase formation \citep{Loomis2015,Oberg2017}, but this could be complemented by an inward drift of ices with H$_2$CO already formed at slightly larger radii \citep{Pegues2020}.

In the outer disk ($\gtrsim$$150$~au), CO is expected to be abundant on ices, and therefore, the reaction represented by Equation~\ref{eq:CO-hydrogenation} could become more efficient. In addition, the shielding from UV radiation is expected to decrease toward the outskirts of the disk, as suggested by previous DCO$^{+}$ observations toward HD 163296 with a similar radial distribution \citep{Salinas2017,Salinas2018}. Thus, the photodesorption of H$_2$CO \citep{Carney2017,Oberg2017} and some CO \citep{Oberg2015b,Huang2016} from grains could increase, which could explain the bump of $N_{\mathrm{T}}$ observed close to $R_{\mathrm{edge}}$. However, the above can also be associated with an enhancement in the production of radicals and atomic O due to the photodissociation of CO in the upper layers, which would favor the gas-phase formation of H$_2$CO in the outer disk \citep{Carney2017}.

Regarding the central depression in the integrated emission, although it is apparently correlated to a decrease in the column density, the results should be treated with caution as we are on a region that cannot be resolved with the current angular resolution, and the quality of the fit is not as good. Alternative explanations are the presence of optically thick dust \citep[e.g.,][]{Carney2017} or the use of lines with low upper-state energies that do not trace the warmer gas in the inner disk \citep[e.g.,][]{Guzman2021}. Since the lines detected in this work with high $E_u$ values also show an inner hole, as well as CO isotopologues \citep{Law2021III}, the scenario of a drop in the emission due to an optically thick continuum seems more likely.

In contrast to the $N_\mathrm{T}$ and OPR profiles, $T_{\mathrm{ex}}$ does not present significant radial substructures, which could imply the dominance of a single formation mechanism throughout the disk, similar to was found in TW~Hya \citep[][]{TvS2021}. However, the temperature gradient in the inner disk is steeper than in the outer disk, rapidly decreasing from $\sim$$50$~K to values around the disk-averaged $T_{\mathrm{ex}} \sim 20$~K, and then decreasing gently at larger radii. This distinction could support the previously suggested two-component H$_2$CO model, with a warm inner region with predominant gas-phase formation and a cold component beyond the CO snowline with contribution from both formation pathways \citep{Loomis2015,Carney2017,Oberg2017,Pegues2020}. 

\cite{Qi2015} estimated a CO freeze-out midplane temperature of $\sim$$25$~K for HD 163296, which is a bit smaller than our derived value of $T_{\mathrm{ex}} \sim 30$~K at $\sim$$75$~au (the location of the CO snowline). Instead, typical CO freeze-out temperatures \citep[$T_{\mathrm{kin}}\lesssim25$~K,][]{Oberg2015b,Qi2015,Pinte2018} are reached in the outer disk, close to $R_{\mathrm{edge}}$, where we expect to have more H$_2$CO from dust grains. The above is a consequence that disk temperature depends on the radius and the vertical height. Thus, even if the H$_2$CO emission comes from a radius outside the CO snowline, this same emission may have originated in a region where CO is not frozen out, emphasizing the importance of complementing our radial analysis with constraints on the vertical height.

\subsubsection{Vertical Distribution of H$_2$CO} \label{subsubsec:Vertical_distrib}

We complemented our radial temperature measurements with the vertical distribution of the emission obtained with \code{ALFAHOR}, as shown in Figure~\ref{fig:vertical_height}. For consistency, we compared our vertical and radial constraints with predictions of the temperature structure in HD 163296 (see background maps in Figure~\ref{fig:vertical_height}). For the inner disk, $T_{\mathrm{ex}}$ values are more consistent with the empirical temperature structure (see left panel in Figure~\ref{fig:vertical_height}), while in the outer disk, $T_{\mathrm{ex}}$ resembles that predicted from thermochemical models (see right panel in Figure~\ref{fig:vertical_height}). These disparities align with the fact that thermochemical models can underpredict the temperature deduced from observations in the inner disk because they do not include mechanisms like accretion heating or CO depletion \citep{Zhang2021}. Moreover, in the outer disk, temperature predictions from CO observations could be more uncertain as the emission from optically thin CO isotopologues is less extended \citep{Law2021IV}. 

Empirical measurements of the vertical height and predictions of the temperature structure agree that H$_2$CO is mostly tracing a cold region relatively close to the midplane ($z/r \gtrsim 0.1$). The intersection of the CO snow surface ($T_{\mathrm{kin}} = 25$~K) with the $z/r = 0.1$ surface is expected at $100-150$~au, depending on the assumed temperature structure (see dash-dotted lines in Figure~\ref{fig:vertical_height}). This spatial feature is consistent with our delineation of the inner and outer disk, based on the column density profile, and could explain why the gas temperature traced by H$_2$CO has different behavior in these two regions, supporting different H$_2$CO formation pathways.

Interestingly, the four analyzed transitions present modulations on the computed vertical height at similar positions as the integrated line intensity. \cite{Paneque-Carreno2023} resolved a similar structure in HD 163296 using the $3_{03}-2_{02}$ line but with higher angular resolution. They discarded the possibility of an effect of the line emission morphology in determining the vertical structure, linking it to actual changes in the vertical location of the H$_2$CO emitting region. The association of the local minimum heights with H$_2$CO emission rings (at $230$ and $378$~au, see Figure~\ref{fig:excitation-conds}) could be related to an increased penetration of UV photons in the outer disk that would enhance the formation of H$_2$CO. Actually, vertical heights predicted in this work are near those layers with active photochemistry traced by HCN and C$_2$H \citep[e.g.,][]{Law2021IV,Paneque-Carreno2023}. Still, the above would be consistent with the grain-surface chemistry (photodesorption of H$_2$CO), but also the cold gas-phase chemistry due to increased production of radicals.

\subsubsection{Ortho-to-para Ratio of H$_2$CO} \label{subsubsec:OPR}

The OPR has been proposed as an indicator of the conditions under which H$_2$CO is formed, based on its relation with the spin temperature \citep[$T_{\mathrm{spin}}$, e.g.,][]{Guzman2018}. Owing to the large timescales involved in the ortho-para nuclear-spin conversion by nonreactive collisions in the gas phase, $T_{\mathrm{spin}}$ is expected to resemble the formation temperature of H$_2$CO \citep{Tudorie2006}. According to the Boltzmann distribution, higher OPR values are associated with higher spin temperatures \citep[see][and references therein]{Guzman2018}. Hence, a statistical value of $3.0$ was initially expected for the gas-phase chemistry, while lower values were thought to result when H$_2$CO forms on the surface of dust grains \citep{Kahane1984,Guzman2011}. Nevertheless, the latter began to be questioned by laboratory experiments where water desorption from ice resulted in an OPR consistent with $3.0$ \citep{Hama2016,Hama2018}. Recent results suggest that the same behavior could occur for H$_2$CO desorption \citep{Yocum2023}, and previous work has discussed the possibility of low OPR values being associated with a cold gas-phase formation from radicals produced at low temperatures \citep{Guzman2018,TvS2021}. It is, therefore, unclear how OPRs are related to the formation pathways of molecules.

From our disk-averaged and resolved analysis, we derived a relatively constant value of OPR$\sim2.7$ across the disk, which could still be consistent with 3.0, considering the uncertainties. Similar to the integrated intensity and the vertical height, the OPR seems to have some radial modulations that could be related to changes in the chemistry\footnote{We cannot discard the possibility that the tentative variations observed in the OPR profile could be due to slight deviations from the assumptions made to derive $N_\mathrm{T}$ (e.g., LTE emission, same $T_{\mathrm{ex}}$ for both isomers and the same emitting region for all the lines).}. The lowest OPR values are found close to the H$_2$CO rings and at lower disk altitudes; thus, they are likely associated with a rise in UV penetration as well. We also extracted the $T_{\mathrm{spin}}$ radial profile from the resolved OPR values (see Fig.~\ref{fig:Tspin_profile}), and found a coincidence between
$T_{\mathrm{spin}}$ and $T_{\mathrm{ex}}$ values, mainly in the outer disk, which could indicate that the formation temperature of H$_2$CO is preserved. However, laboratory experiments have shown that $T_{\mathrm{spin}}$ does not correspond to the ice formation temperature when H$_2$CO is formed on ices by the UV photolysis of CH$_3$OH \citep{Yocum2023}. Still, it is also possible that $T_{\mathrm{spin}}$ could reflect the formation temperature if H$_2$CO forms more quiescently through the hydrogenation of CO ices.

Therefore, our results are consistent with both the cold gas-phase and grain-surface formation scenarios in HD 163296. Indeed, more in-depth studies of dust-grain-related processes are needed to discern if grain-surface chemistry is critical for H$_2$CO formation in HD 163296, as well as its role in setting the OPR value. Since gas-phase products may also contribute to the formation of H$_2$CO on grains \citep{TvS2021}, chemical models considering both mechanisms are an alternative but, unfortunately, the physical parameters involved in grain-surface chemistry are poorly constrained \citep[][]{Guzman2013}. Modeling only gas-phase chemistry \citep[e.g.,][]{Loomis2015,Loomis2018} could be a more straightforward approach, but it is beyond the scope of this work.

\subsection{H$_2$CO in T Tauri versus Herbig Ae/Be disks}

Thanks to ALMA, H$_2$CO has been currently resolved in $\sim$20 protoplanetary disks \citep[see][and references therein]{Pegues2020,TvS2021}. From this sample, only five correspond to Herbig Ae/Be disks: the cold disks HD 163296 (this work) and MWC 480 \citep{Pegues2020,Guzman2021}, and the warm disks IRS 48 \citep{vanderMarel2021,Booth2024II}, HD 100546 \citep{Booth2021,Booth2024I}, and HD 169142 \citep{Booth2023}.

MWC 480 has similar $N_{\mathrm{T}}$ and $T_{\mathrm{ex}}$ values to HD 163296, while higher $T_{\mathrm{ex}}$ values and radial substructures of $N_{\mathrm{T}}$ correlated with the dust continuum characterize IRS 48, HD 100546, and HD 169142. These differences reflect how H$_2$CO chemistry acts under different physical conditions since, typically, warm Herbig Ae/Be disks are characterized by dust cavities, which prevent the in situ formation of CO ices, and by the presence of CH$_3$OH detections \citep{Booth2023,Booth2024II,Booth2024I}. In these cases, H$_2$CO formation is probably dominated by warm gas-phase chemistry and/or thermal desorption from inherited ices, as suggested by the low OPR of $1.2$ deduced in IRS 48 \citep{vanderMarel2021}.

Instead, the morphology of $N_{\mathrm{T}}$ and the disk-averaged $T_{\mathrm{ex}}$ of H$_2$CO in cold Herbig Ae/Be disks are more similar to those observed in T Tauri disks \citep[e.g.,][]{Pegues2020}. Still, the $N_{\mathrm{T}}$ values of cold Herbig Ae/Be disks are systematically lower than those of T Tauri disks, which has been explained as a consequence of the higher temperatures expected around more massive stars that limit the CO freeze-out \citep{Pegues2020}, in line with the two-component H$_2$CO model discussed in Section~\ref{subsubsec:Vertical_distrib}. To verify this possibility of similar chemistry, we can analyze the radial behavior of the OPR, where the only comparison that can be made is with the T Tauri disk TW Hya. Interestingly, unlike the relatively constant OPR observed in HD 163296, the OPR profile of TW Hya has an apparent decreasing behavior \citep{TvS2021}, thus suggesting a different formation process. Further experiments of the OPR in H$_2$CO ices and more determinations of the OPR around both T Tauri and Herbig Ae/Be protoplanetary disks are needed to confirm this scenario.

TW Hya is a relatively peculiar source. It is relatively old compared to other T Tauri stars, and it is the only T Tauri disk with known detections of CH$_3$OH \citep{Walsh2016}. Also, in terms of H$_2$CO, it has a relatively high and nearly constant excitation temperature ($T_{\mathrm{ex}} \sim 30 - 40$~K) and a different column density morphology (no rings in the outer disk), in comparison with other T Tauri disks and HD 163296. Moreover, constraints on the vertical structure have placed the emitting surface of H$_2$CO in TW Hya at higher heights \citep[$z/r \gtrsim 0.25$,][]{Oberg2017,TvS2021} than in other disks with empirical determinations \citep[$z/r \sim 0.1$,][]{Paneque-Carreno2023}. These differences align with the fact that TW Hya is one of the few disks where H$_2$CO is proposed to be predominantly formed by gas phase rather than a combination of both formation pathways \citep{TvS2021}. Observations toward other disks are needed to better understand the dominant formation pathway of H$_2$CO.

Curiously, even though it has no efficient gas-phase formation pathways, CH$_3$OH is relatively abundant in TW Hya (CH$_3$OH/H$_2$CO $=1.27$) but not so in HD 163296 (CH$_3$OH/H$_2$CO $< 0.24$), according to the upper limits derived by \cite{Carney2019}. One possibility to address this counterintuitive behavior could be a difference in the UV radiation field. UV photons can photodesorb intact CH$_3$OH from ices, but also photofragments of H$_2$CO \citep{Guzman2013,Bertin2016,Yocum2023}. Because massive Herbig Ae/Be disks have stronger UV fluxes than T Tauri disks \citep{Walsh2015}, maybe CH$_3$OH tends to be more fragmented for sources like HD 163296, thus contributing to H$_2$CO emission instead of the gas-phase CH$_3$OH, as supposed in TW Hya. 

Furthermore, laboratory experiments have demonstrated that X-ray photodesorption is also an efficient mechanism to release CH$_3$OH into the gas phase, even more than UV photodesorption \citep{Basalgete2021I,Basalgete2021II}. The above could also contribute to differences in the CH$_3$OH/H$_2$CO ratio since T Tauri sources have larger X-ray luminosities than Herbig Ae/Be \citep{Ryspaeva2023}, and therefore, the X-ray-induced chemistry is predicted to be more significant \citep{Walsh2015}. More complementary detections of H$_2$CO and upper limits for CH$_3$OH (if not detected) in protoplanetary disks are needed to clarify this behavior.

\section{Conclusions}\label{sec:Conclusions}

We have presented the analysis of multiple spectrally and spatially resolved transitions for the para and ortho isomers of H$_2$CO toward the Herbig Ae disk around HD 163296. The main conclusions can be summarized as follows:

\begin{enumerate}
    \item The line emission, extended up to $\sim$$450$~au, is characterized by the presence of a central depression and a global peak at $\sim$$60$~au in the inner disk, whereas in the outer disk, two ring-like structures are located at $\sim$$250$ and $\sim$$380$~au.
    \item According to LTE predictions, we determine the column density of H$_2$CO as a function of disk radius, and for the first time, we resolve the excitation temperature and the ortho-to-para ratio of H$_2$CO in HD 163296.
    \item The column density values range from $\sim 2.0\times 10^{13}$ to $\sim 1.3\times 10^{12}$~cm$^{-2}$, showing a peak in the inner disk and a plateau in the outer disk, likely related to the positions of the CO snowline and the dust millimeter edge, respectively.
    \item The excitation temperature, which is expected to trace the temperature of the emitting gas, decreases smoothly from $\sim$$50$ to $\sim$$10$~K throughout the disk. Still, the slope of the profile in the inner disk is steeper than in the outer disk.
    \item The temperature profile was complemented with empirical measurements of the vertical height of H$_2$CO emission. By comparing with predictions of the temperature structure in HD 163296, we find that H$_2$CO is tracing a cold region near the midplane ($z/r \gtrsim 0.1$).
    \item An ortho-to-para ratio close to 2.7 remains relatively constant across the disk, which is still consistent with the statistical value of 3.0, considering the uncertainties. However, the OPR profile presents radial modulations, where the lowest OPR values are coincident with H$_2$CO emission rings and lower vertical heights.
    \item Constraints from the excitation conditions and the ortho-to-para ratio are consistent with the formation of H$_2$CO by both gas-phase and by grain-surface chemistry, especially in the outer disk, where there is an apparent increase in the UV penetration that would favor the photodesorption of H$_2$CO from dust grains, but also the formation of radicals due to CO photodissociation.
    \item Our results suggest that the formation of H$_2$CO in relatively cold Herbig Ae/Be disks (e.g., HD 163296 and MWC 480) is different from that in their warmer analogs characterized by dust cavities and CH$_3$OH detections (e.g., IRS 48, HD 100546, and HD 169142). In contrast, the morphology of $N_\mathrm{T}$ and the disk-averaged $T_{\mathrm{ex}}$ of H$_2$CO in HD 163296 shares similarities with the results in T Tauri disks.
    \item Regarding the OPR, the only source where a comparison can be made is TW Hya, a T Tauri disk where H$_2$CO is thought to be predominantly formed by gas-phase chemistry. Although the OPR values are consistent with $3.0$ in the inner disk of both sources, TW Hya exhibits a monotonic decrease at the outer radius that is not seen for HD 163296, suggesting differences in the chemistry. Dissimilarities in the column density and excitation temperature profiles also support the above. Further multiple-line observations toward more T Tauri and Herbig Ae/Be sources are needed to disentangle the dominant formation pathway of H$_2$CO in protoplanetary disks.
\end{enumerate}

\section*{Acknowledgments}

This paper makes use of the following ALMA data: ADS/JAO.ALMA\#2016.1.00884.S and ADS/JAO.ALMA\#2018.1.01055.L. ALMA is a partnership of ESO (representing its member states), NSF (USA) and NINS (Japan), together with NRC (Canada), MOST and ASIAA (Taiwan), and KASI (Republic of Korea), in cooperation with the Republic of Chile. The Joint ALMA Observatory is operated by ESO, AUI/NRAO and NAOJ. The National Radio Astronomy Observatory is a facility of the National Science Foundation operated under cooperative agreement by Associated Universities, Inc. 

We thank the anonymous reviewer for providing useful comments and suggestions. C.H.-V. acknowledges support from the National Agency for Research and Development (ANID) -- Scholarship Program through the Doctorado Nacional grant No. 2021-21212409. 
V.V.G gratefully acknowledges support from FONDECYT Regular 1221352, and ANID CATA-BASAL project FB210003.
L.I.C. acknowledges support from NASA ATP 80NSSC20K0529. L.I.C. also acknowledges support from NSF grant no. AST-2205698, the David and Lucille Packard Foundation, and the Research Corporation for Scientific Advancement Cottrell Scholar Award.

\software{\code{ALFAHOR} \citep{Paneque-Carreno2023}, Astropy \citep{Astropy2013,Astropy2018,Astropy2022}, \code{bettermoments} \citep{Teague2018bettermoments}, \code{CASA} \citep{McMullin2007,CASA2022}, \code{emcee} \citep{Foreman-Mackey2013}, \code{GoFish} \citep{Teague2019GoFish}, Matplotlib \citep{Hunter2007}, NumPy \citep{vanderWalt2011}, SciPy \citep{Virtanen2020}.}

\clearpage
\appendix

\vspace*{-0.5cm}
\section{Observational details}\label{sec:Appendix_imaging}

Table~\ref{table:imaging_parameters} summarizes the imaging parameters to obtain a common circularized beam of $0\farcs6 \times 0\farcs6$ for all transitions used in this work and the $\epsilon$ parameter to account for the \textit{JvM effect}.

\begin{deluxetable}{c c c c}[h]
\tabletypesize{\footnotesize}
\tablewidth{0pt} 
\tablenum{A1}
\label{table:imaging_parameters}
\tablecaption{Imaging parameters}
\tablehead{
\colhead{H$_2$CO Line} & \colhead{Robust} & \colhead{UV taper} & \colhead{JvM $\epsilon$} \\
\colhead{} & \colhead{} & \colhead{($^{\prime\prime}$)} & \colhead{}}
\startdata
\multicolumn{4}{c}{ALMA band 3} \\
o-H$_2$CO $6_{15}-6_{16}$ & $0.5$ & $0.556 \times 0.433, -2.287$ & $0.80$ \\ 
\hline 
\multicolumn{4}{c}{ALMA band 6} \\
p-H$_2$CO $3_{03}-2_{02}$ & $0.5$ & $0.520 \times 0.571, 71.583$ & $0.92$ \\
o-H$_2$CO $3_{12}-2_{11}$ & $0.5$ & $0.146 \times 0.417, 79.759$ & $0.59$\\
\hline 
\multicolumn{4}{c}{ALMA band 7} \\
p-H$_2$CO $4_{04}-3_{03}$ & $0.0$ & $0.167 \times 0.476, 80.301$ & $0.98$ \\
p-H$_2$CO $4_{23}-3_{22}$ & $0.0$ & $0.194 \times 0.473, 85.875$ & $0.99$ \\
p-H$_2$CO $4_{22}-3_{21}$ & $0.0$ & $0.126 \times 0.489, 73.735$ & $0.99$\\
o-H$_2$CO $4_{13}-3_{12}$ & $0.25$ & $0.001 \times 0.435, 83.675$ & $0.97$ \\
\enddata
\end{deluxetable}

\vspace*{-1.9cm}
\section{Voigt profile line modeling}\label{sec:Appendix_voigt}

\begin{figure*}[h!]
\centering
\includegraphics[width=0.95\linewidth]{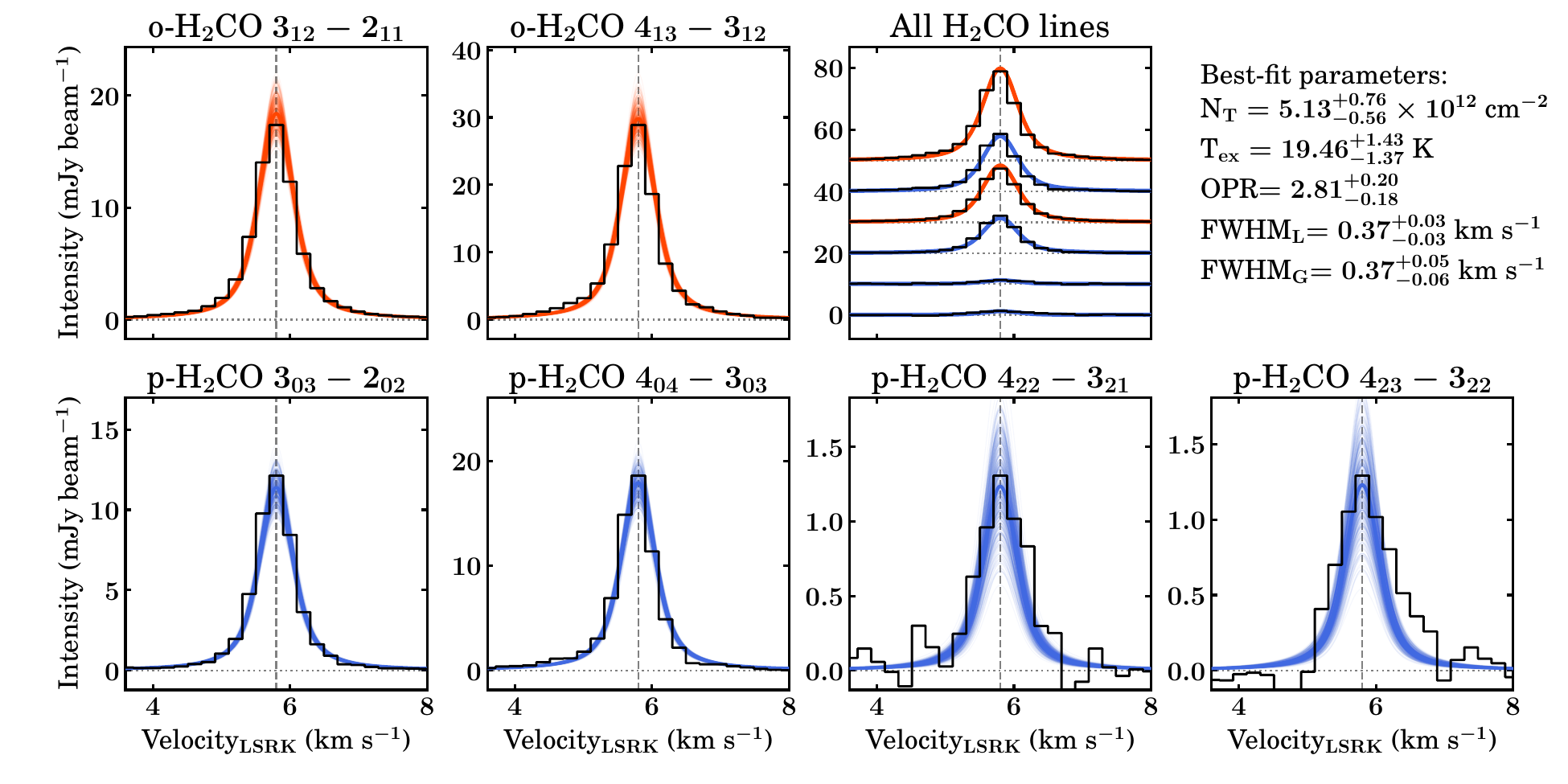}
\caption{Same as Figure~\ref{fig:fitted_disk-averaged_spectra}, but using Voigt profiles to model the disk-averaged emission of H$_2$CO. The FWHM$_\mathrm{L}$ and FHWM$_\mathrm{G}$ values represent the Lorentz and Gaussian line width, respectively.}
\label{fig:fitted_disk-averaged_spectra_voigt}
\end{figure*}

Following the analysis described in Section~\ref{subsec:excitation-conditions}, we model the disk-averaged emission using Voigt profiles instead of Gaussian to better characterize the line wings. The retrieved disk-averaged spectra and the best-fit parameters are shown in Figure~\ref{fig:fitted_disk-averaged_spectra_voigt}, where the best-fit models are overlaid with different colors for each H$_2$CO isomer. A similar approach was applied in TW Hya for $^{12}$CO observations likely affected by line pressure broadening \citep{Yoshida2022}, which enabled the estimation of its gas surface density. Therefore, the presence of non-Gaussian profiles in HD 163296 could have some implications in constraining its physical conditions, but this is beyond the scope of this work.

\newpage

\section{Optical depth radial profiles}\label{sec:Appendix_tauprofiles}

\begin{figure*}[h!]
\centering
\includegraphics[width=0.5\linewidth]{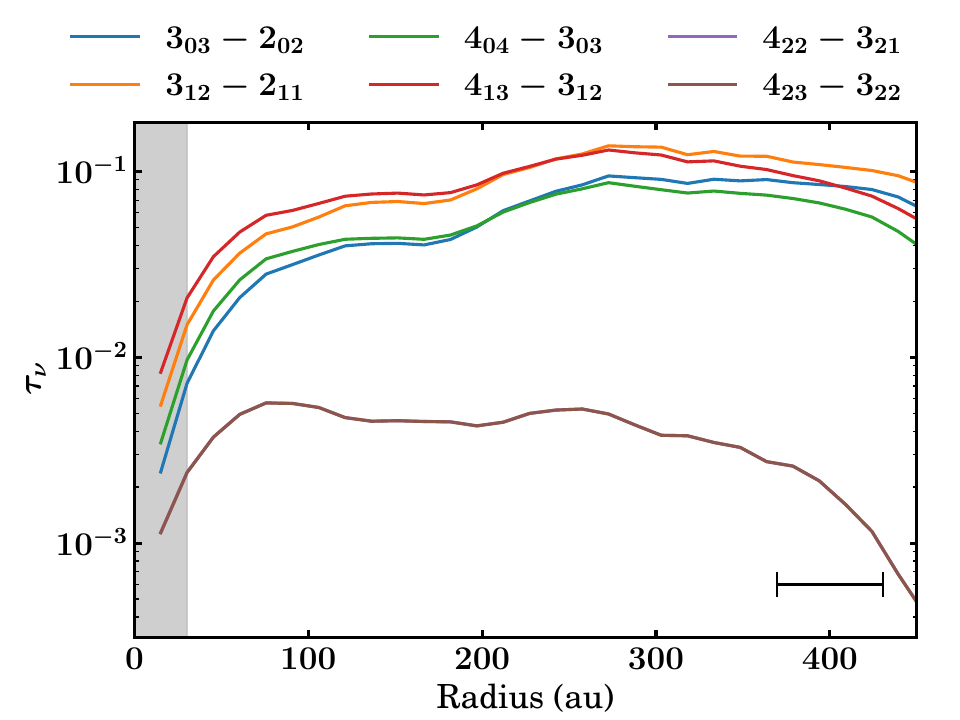}
\caption{Radial profiles of the line optical depth of H$_2$CO. The beam size is represented by the horizontal black bar, while the vertical gray-shaded region denotes an extent
equivalent to half of the beam size, where values should be treated with caution.}
\label{fig:tau_profiles}
\end{figure*}

Figure~\ref{fig:tau_profiles} shows the deprojected radial profiles of the line optical depth for all H$_2$CO transitions used in this work, resulting from the analysis described in Section~\ref{subsubsec:Resolved_analysis}. Each color represents a different transition. The two weakest p-H$_2$CO lines ($4_{22}-3_{21}$ and $4_{23}-3_{22}$) have very similar values for $\tau_{\nu}$; hence, the two curves are overlapping.

\section{H$_2$CO formation temperature}\label{sec:Appendix_Tspin}

\begin{figure*}[h!]
\centering
\includegraphics[width=0.5\linewidth]{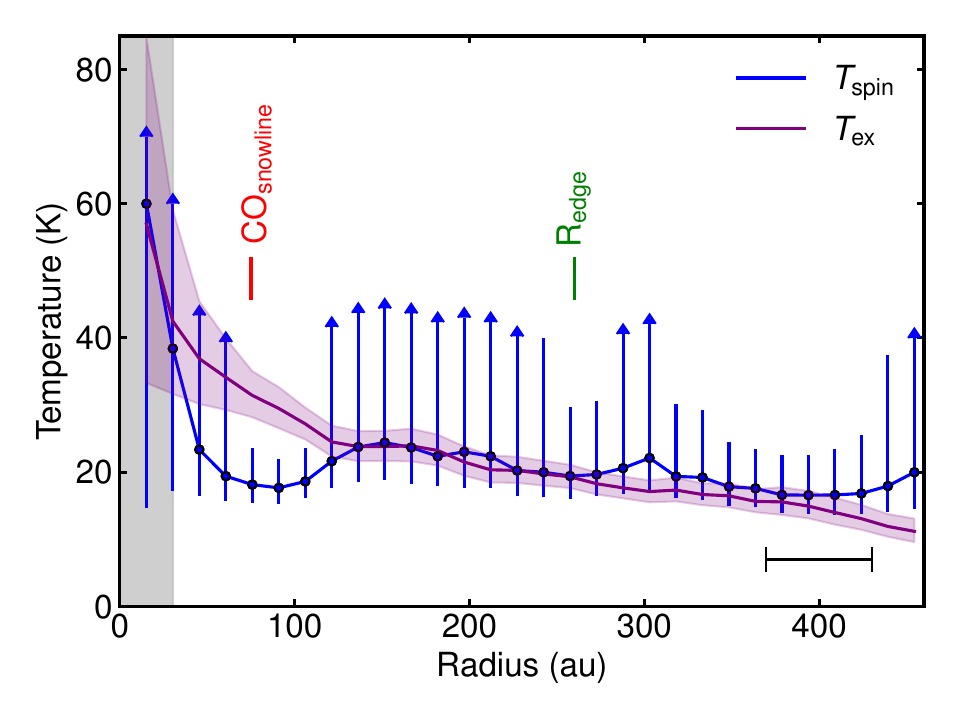}
\caption{Radial profiles of the excitation temperature (purple curve) and spin temperature (blue curve). Vertical lines depict the CO snowline (red) and the dust millimeter edge (green). The horizontal black bar and the vertical gray-shaded region are the same as in Figure~\ref{fig:tau_profiles}.}
\label{fig:Tspin_profile}
\end{figure*}

Figure~\ref{fig:Tspin_profile} shows the deprojected radial profiles of the excitation temperature (see middle panel in Figure~\ref{fig:excitation-conds}) and the spin temperature derived from the radially resolved OPR values (see bottom panel in Figure~\ref{fig:excitation-conds}). As a consequence of the asymptotic behavior of the OPR as a function of $T_\mathrm{spin}$ \cite[see Figure~7 in][]{Guzman2018}, the arrows represent lower limits in the upper uncertainties of $T_\mathrm{spin}$ for those radii where the OPR upper uncertainties are consistent with $\mathrm{OPR} \geq 3.0$.

\bibliography{sample631}{}
\bibliographystyle{aasjournal}

\end{document}